%%%%%%%%%%%%%%%%%%%%%%%%%%%%%%%%%%%%%%%%%%%%%%%%

%harvmac
%\input harvmac

%%%%%%%%%%%%%%%%%%  tex macros for preprints, cm version %%%%%%%%%%%%%%
%                     (P. Ginsparg, last updated 9/91)
%                if confused, type `b' in response to query
%
%---------------------------------------------------------------------%
%% site dependent options:
%% \unredoffs and \redoffs define horizontal and vertical offsets
%% respectively for unreduced and reduced modes. \speclscape defines
%% the \special{} call that sets printer to landscape (sideways) mode.
%% from standard set below, leave uncommented as appropriate or redefine
%
%%% next 400dpi
%\def\unredoffs{} \def\redoffs{\voffset=-.31truein\hoffset=-.48truein}
%\def\speclscape{\special{landscape}}
%
%%% apple lw
\def\unredoffs{} \def\redoffs{\voffset=-.31truein\hoffset=-.59truein}
\def\speclscape{\special{ps: landscape}}
%
%%% qms lasergrafix:
%\def\unredoffs{} \def\redoffs{\voffset=-.4truein\hoffset=.125truein}
%\def\speclscape{\special{qms: landscape}}
%
%%% saclay A4 paper:
%\def\unredoffs{\hoffset-.14truein\voffset-.2truein}
%\def\redoffs{\voffset=-.45truein\hoffset=-.21truein}
%\def\speclscape{\special{landscape}}
%
%---------------------------------------------------------------------%
%
\newbox\leftpage \newdimen\fullhsize \newdimen\hstitle \newdimen\hsbody
\tolerance=1000\hfuzz=2pt
\catcode`\@=11 % This allows us to modify PLAIN macros.
\def\bigans{b }
\def\answ{b }

%\message{ big or little (b/l)? }\read-1 to\answ
%
\ifx\answ\bigans\message{(This will come out unreduced.}
\magnification=1200\unredoffs\baselineskip=16pt plus 2pt minus 1pt
\hsbody=\hsize \hstitle=\hsize %take default values for unreduced format
\else\message{(This will be reduced.} \let\l@r=L
\magnification=1000\baselineskip=16pt plus 2pt minus 1pt \vsize=7truein
\redoffs \hstitle=8truein\hsbody=4.75truein\fullhsize=10truein\hsize=\hsbody
\output={\ifnum\pageno=0 %%% This is the HUTP version
  \shipout\vbox{\speclscape{\hsize\fullhsize\makeheadline}
    \hbox to \fullhsize{\hfill\pagebody\hfill}}\advancepageno
  \else
  \almostshipout{\leftline{\vbox{\pagebody\makefootline}}}\advancepageno
  \fi}
\def\almostshipout#1{\if L\l@r \count1=1 \message{[\the\count0.\the\count1]}
      \global\setbox\leftpage=#1 \global\let\l@r=R
 \else \count1=2
  \shipout\vbox{\speclscape{\hsize\fullhsize\makeheadline}
      \hbox to\fullhsize{\box\leftpage\hfil#1}}  \global\let\l@r=L\fi}
\fi
%---------------------------------------------------------------------
%
\newcount\yearltd\yearltd=\year\advance\yearltd by -1900

\def\Title#1#2{\nopagenumbers\abstractfont\hsize=\hstitle\rightline{#1}%
\vskip 1in\centerline{\titlefont #2}\abstractfont\vskip .5in\pageno=0}
\def\Date#1{\vfill\leftline{#1}\tenpoint\supereject\global\hsize=\hsbody%
\footline={\hss\tenrm\folio\hss}}%  restores pagenumbers
%
%       use following instead of \Date on the preliminary draft,
%       puts date/time on each page in big mode, writes labels in margins

\def\draftmode{\message{ DRAFTMODE }\def\draftdate{{\rm preliminary draft:
\number\month/\number\day/\number\yearltd\ \ \hourmin}}%
\headline={\hfil\draftdate}\writelabels\baselineskip=20pt plus 2pt minus 2pt
 {\count255=\time\divide\count255 by 60 \xdef\hourmin{\number\count255}
  \multiply\count255 by-60\advance\count255 by\time
  \xdef\hourmin{\hourmin:\ifnum\count255<10 0\fi\the\count255}}}
%       use \nolabels to get rid of eqn, ref, and fig labels in draft mode
\def\nolabels{\def\wrlabeL##1{}\def\eqlabeL##1{}\def\reflabeL##1{}}
\def\writelabels{\def\wrlabeL##1{\leavevmode\vadjust{\rlap{\smash%
{\line{{\escapechar=` \hfill\rlap{\sevenrm\hskip.03in\string##1}}}}}}}%
\def\eqlabeL##1{{\escapechar-1\rlap{\sevenrm\hskip.05in\string##1}}}%
\def\reflabeL##1{\noexpand\llap{\noexpand\sevenrm\string\string\string##1}}}
\nolabels
%
% tagged sec numbers
\global\newcount\secno \global\secno=0
\global\newcount\meqno \global\meqno=1
\def\newsec#1{\global\advance\secno by1\message{(\the\secno. #1)}
%\ifx\answ\bigans \vfill\eject \else \bigbreak\bigskip \fi  %if desired
\global\subsecno=0\eqnres@t\noindent{\bf\the\secno. #1}
\writetoca{{\secsym} {#1}}\par\nobreak\medskip\nobreak}
\def\eqnres@t{\xdef\secsym{\the\secno.}\global\meqno=1\bigbreak\bigskip}
\def\sequentialequations{\def\eqnres@t{\bigbreak}}\xdef\secsym{}
\global\newcount\subsecno \global\subsecno=0
\def\subsec#1{\global\advance\subsecno by1\message{(\secsym\the\subsecno. #1)}
\ifnum\lastpenalty>9000\else\bigbreak\fi
\noindent{\it\secsym\the\subsecno. #1}\writetoca{\string\quad
{\secsym\the\subsecno.} {#1}}\par\nobreak\medskip\nobreak}
\def\appendix#1#2{\global\meqno=1\global\subsecno=0\xdef\secsym{\hbox{#1.}}
\bigbreak\bigskip\noindent{\bf Appendix #1. #2}\message{(#1. #2)}
\writetoca{Appendix {#1.} {#2}}\par\nobreak\medskip\nobreak}
%
%       \eqn\label{a+b=c}   gives displayed equation, numbered
%               consecutively within sections.
%     \eqnn and \eqna define labels in advance (of eqalign?)
%
\def\eqnn#1{\xdef #1{(\secsym\the\meqno)}\writedef{#1\leftbracket#1}%
\global\advance\meqno by1\wrlabeL#1}
\def\eqna#1{\xdef #1##1{\hbox{$(\secsym\the\meqno##1)$}}
\writedef{#1\numbersign1\leftbracket#1{\numbersign1}}%
\global\advance\meqno by1\wrlabeL{#1$\{\}$}}
\def\eqn#1#2{\xdef #1{(\secsym\the\meqno)}\writedef{#1\leftbracket#1}%
\global\advance\meqno by1$$#2\eqno#1\eqlabeL#1$$}
%
%            footnotes
\newskip\footskip\footskip14pt plus 1pt minus 1pt %sets footnote baselineskip
\def\footnotefont{\ninepoint}\def\f@t#1{\footnotefont #1\@foot}
\def\f@@t{\baselineskip\footskip\bgroup\footnotefont\aftergroup\@foot\let\next}
\setbox\strutbox=\hbox{\vrule height9.5pt depth4.5pt width0pt}
\global\newcount\ftno \global\ftno=0
\def\foot{\global\advance\ftno by1\footnote{$^{\the\ftno}$}}
%
%say \footend to put footnotes at end
%will cause problems if \ref used inside \foot, instead use \nref before
\newwrite\ftfile
\def\footend{\def\foot{\global\advance\ftno by1\chardef\wfile=\ftfile
$^{\the\ftno}$\ifnum\ftno=1\immediate\openout\ftfile=foots.tmp\fi%
\immediate\write\ftfile{\noexpand\smallskip%
\noexpand\item{f\the\ftno:\ }\pctsign}\findarg}%
\def\footatend{\vfill\eject\immediate\closeout\ftfile{\parindent=20pt
\centerline{\bf Footnotes}\nobreak\bigskip\input foots.tmp }}}
\def\footatend{}
%
%     \ref\label{text}
% generates a number, assigns it to \label, generates an entry.
% To list the refs on a separate page,  \listrefs
%
\global\newcount\refno \global\refno=1
\newwrite\rfile
\def\ref{[\the\refno]\nref}
\def\nref#1{\xdef#1{[\the\refno]}\writedef{#1\leftbracket#1}%
\ifnum\refno=1\immediate\openout\rfile=refs.tmp\fi
\global\advance\refno by1\chardef\wfile=\rfile\immediate
\write\rfile{\noexpand\item{#1\ }\reflabeL{#1\hskip.31in}\pctsign}\findarg}
%   horrible hack to sidestep tex \write limitation
\def\findarg#1#{\begingroup\obeylines\newlinechar=`\^^M\pass@rg}
{\obeylines\gdef\pass@rg#1{\writ@line\relax #1^^M\hbox{}^^M}%
\gdef\writ@line#1^^M{\expandafter\toks0\expandafter{\striprel@x #1}%
\edef\next{\the\toks0}\ifx\next\em@rk\let\next=\endgroup\else\ifx\next\empty%
\else\immediate\write\wfile{\the\toks0}\fi\let\next=\writ@line\fi\next\relax}}
\def\striprel@x#1{} \def\em@rk{\hbox{}}
\def\lref{\begingroup\obeylines\lr@f}
\def\lr@f#1#2{\gdef#1{\ref#1{#2}}\endgroup\unskip}

\def\addref#1{\immediate\write\rfile{\noexpand\item{}#1}} %now unnecessary
\def\startrefs#1{\immediate\openout\rfile=refs.tmp\refno=#1}
\def\xref{\expandafter\xr@f}\def\xr@f[#1]{#1}
\def\refs#1{\count255=1[\r@fs #1{\hbox{}}]}
\def\r@fs#1{\ifx\und@fined#1\message{reflabel \string#1 is undefined.}%
\nref#1{need to supply reference \string#1.}\fi%
\vphantom{\hphantom{#1}}\edef\next{#1}\ifx\next\em@rk\def\next{}%
\else\ifx\next#1\ifodd\count255\relax\xref#1\count255=0\fi%
\else#1\count255=1\fi\let\next=\r@fs\fi\next}
%

%
% this is ugly, but moore insists
\newwrite\ffile\global\newcount\figno \global\figno=1
\def\fig{fig.~\the\figno\nfig}
\def\nfig#1{\xdef#1{fig.~\the\figno}%
\writedef{#1\leftbracket fig.\noexpand~\the\figno}%
\ifnum\figno=1\immediate\openout\ffile=figs.tmp\fi\chardef\wfile=\ffile%
\immediate\write\ffile{\noexpand\medskip\noexpand\item{Fig.\ \the\figno. }
\reflabeL{#1\hskip.55in}\pctsign}\global\advance\figno by1\findarg}
\def\xfig{\expandafter\xf@g}\def\xf@g fig.\penalty\@M\ {}
\def\figs#1{figs.~\f@gs #1{\hbox{}}}
\def\f@gs#1{\edef\next{#1}\ifx\next\em@rk\def\next{}\else
\ifx\next#1\xfig #1\else#1\fi\let\next=\f@gs\fi\next}
\newwrite\lfile
{\escapechar-1\xdef\pctsign{\string\%}\xdef\leftbracket{\string\{}
\xdef\rightbracket{\string\}}\xdef\numbersign{\string\#}}

\def\writestop{\def\writestoppt{\immediate\write\lfile{\string\pageno%
\the\pageno\string\startrefs\leftbracket\the\refno\rightbracket%
\string\def\string\secsym\leftbracket\secsym\rightbracket%
\string\secno\the\secno\string\meqno\the\meqno}\immediate\closeout\lfile}}
\def\writestoppt{}\def\writedef#1{}
\def\seclab#1{\xdef #1{\the\secno}\writedef{#1\leftbracket#1}\wrlabeL{#1=#1}}
\def\subseclab#1{\xdef #1{\secsym\the\subsecno}%
\writedef{#1\leftbracket#1}\wrlabeL{#1=#1}}
\newwrite\tfile \def\writetoca#1{}
\def\leaderfill{\leaders\hbox to 1em{\hss.\hss}\hfill}
%   use this to write file with table of contents
\def\writetoc{\immediate\openout\tfile=toc.tmp
   \def\writetoca##1{{\edef\next{\write\tfile{\noindent ##1
   \string\leaderfill {\noexpand\number\pageno} \par}}\next}}}
%       and this lists table of contents on second pass

%
\catcode`\@=12 % at signs are no longer letters
%
%   Unpleasantness in calling in abstract and title fonts
\edef\tfontsize{\ifx\answ\bigans scaled\magstep3\else scaled\magstep4\fi}
\font\titlerm=cmr10 \tfontsize \font\titlerms=cmr7 \tfontsize
\font\titlermss=cmr5 \tfontsize \font\titlei=cmmi10 \tfontsize
\font\titleis=cmmi7 \tfontsize \font\titleiss=cmmi5 \tfontsize
\font\titlesy=cmsy10 \tfontsize \font\titlesys=cmsy7 \tfontsize
\font\titlesyss=cmsy5 \tfontsize \font\titleit=cmti10 \tfontsize
\skewchar\titlei='177 \skewchar\titleis='177 \skewchar\titleiss='177
\skewchar\titlesy='60 \skewchar\titlesys='60 \skewchar\titlesyss='60
\def\titlefont{\def\rm{\fam0\titlerm}% switch to title font
\textfont0=\titlerm \scriptfont0=\titlerms \scriptscriptfont0=\titlermss
\textfont1=\titlei \scriptfont1=\titleis \scriptscriptfont1=\titleiss
\textfont2=\titlesy \scriptfont2=\titlesys \scriptscriptfont2=\titlesyss
\textfont\itfam=\titleit \def\it{\fam\itfam\titleit}\rm}
 \ifx\answ\bigans\else scaled\magstep1\fi
\ifx\answ\bigans\def\abstractfont{\tenpoint}\else
\font\abssl=cmsl10 scaled \magstep1
\font\absrm=cmr10 scaled\magstep1 \font\absrms=cmr7 scaled\magstep1
\font\absrmss=cmr5 scaled\magstep1 \font\absi=cmmi10 scaled\magstep1
\font\absis=cmmi7 scaled\magstep1 \font\absiss=cmmi5 scaled\magstep1
\font\abssy=cmsy10 scaled\magstep1 \font\abssys=cmsy7 scaled\magstep1
\font\abssyss=cmsy5 scaled\magstep1 \font\absbf=cmbx10 scaled\magstep1
\skewchar\absi='177 \skewchar\absis='177 \skewchar\absiss='177
\skewchar\abssy='60 \skewchar\abssys='60 \skewchar\abssyss='60
\def\abstractfont{\def\rm{\fam0\absrm}% switch to abstract font
\textfont0=\absrm \scriptfont0=\absrms \scriptscriptfont0=\absrmss
\textfont1=\absi \scriptfont1=\absis \scriptscriptfont1=\absiss
\textfont2=\abssy \scriptfont2=\abssys \scriptscriptfont2=\abssyss
\textfont\itfam=\bigit \def\it{\fam\itfam\bigit}\def\footnotefont{\tenpoint}%
\textfont\slfam=\abssl \def\sl{\fam\slfam\abssl}%
\textfont\bffam=\absbf \def\bf{\fam\bffam\absbf}\rm}\fi
\def\tenpoint{\def\rm{\fam0\tenrm}% switch back to 10-point type
\textfont0=\tenrm \scriptfont0=\sevenrm \scriptscriptfont0=\fiverm
\textfont1=\teni  \scriptfont1=\seveni  \scriptscriptfont1=\fivei
\textfont2=\tensy \scriptfont2=\sevensy \scriptscriptfont2=\fivesy
\textfont\itfam=\tenit \def\it{\fam\itfam\tenit}\def\footnotefont{\ninepoint}%
\textfont\bffam=\tenbf \def\bf{\fam\bffam\tenbf}\def\sl{\fam\slfam\tensl}\rm}
\font\ninerm=cmr9 \font\sixrm=cmr6 \font\ninei=cmmi9 \font\sixi=cmmi6
\font\ninesy=cmsy9 \font\sixsy=cmsy6 \font\ninebf=cmbx9
\font\nineit=cmti9 \font\ninesl=cmsl9 \skewchar\ninei='177
\skewchar\sixi='177 \skewchar\ninesy='60 \skewchar\sixsy='60
\def\ninepoint{\def\rm{\fam0\ninerm}% switch to footnote font
\textfont0=\ninerm \scriptfont0=\sixrm \scriptscriptfont0=\fiverm
\textfont1=\ninei \scriptfont1=\sixi \scriptscriptfont1=\fivei
\textfont2=\ninesy \scriptfont2=\sixsy \scriptscriptfont2=\fivesy
\textfont\itfam=\ninei \def\it{\fam\itfam\nineit}\def\sl{\fam\slfam\ninesl}%
\textfont\bffam=\ninebf \def\bf{\fam\bffam\ninebf}\rm}
%
%---------------------------------------------------------------------
%

\hyphenation{anom-aly anom-alies coun-ter-term coun-ter-terms}
\def\inv{^{\raise.15ex\hbox{${\scriptscriptstyle -}$}\kern-.05em 1}}

\def\Dsl{\,\raise.15ex\hbox{/}\mkern-13.5mu D} %this one can be subscripted
\def\dsl{\raise.15ex\hbox{/}\kern-.57em\partial}

\font\bigit=cmti10 scaled \magstep1
 %pound sterling
\def\lspace{\ifx\answ\bigans{}\else\qquad\fi}
\def\lbspace{\ifx\answ\bigans{}\else\hskip-.2in\fi} % $$\lbspace...$$
\def\boxeqn#1{\vcenter{\vbox{\hrule\hbox{\vrule\kern3pt\vbox{\kern3pt
    \hbox{${\displaystyle #1}$}\kern3pt}\kern3pt\vrule}\hrule}}}
\def\mbox#1#2{\vcenter{\hrule \hbox{\vrule height#2in
        \kern#1in \vrule} \hrule}}  %e.g. \mbox{.1}{.1}
%   matters of taste
%\def\tilde{\widetilde} \def\bar{\overline} \def\hat{\widehat}
%
% some sample definitions
  %     curly letters

\def\darr#1{\raise1.5ex\hbox{$\leftrightarrow$}\mkern-16.5mu #1}
 %pound sterling

 %puts a small half in a displayed eqn
\def\roughly#1{\raise.3ex\hbox{$#1$\kern-.75em\lower1ex\hbox{$\sim$}}}

%\draftmode
\let\includefigures=\iftrue
\let\useblackboard=\iftrue
\newfam\black

%Figure Stuff
\includefigures
\message{If you do not have epsf.tex (to include figures),}
\message{change the option at the top of the tex file.}
\input epsf
\def\figin{\epsfcheck\figin}\def\figins{\epsfcheck\figins}
\def\epsfcheck{\ifx\epsfbox\UnDeFiNeD
\message{(NO epsf.tex, FIGURES WILL BE IGNORED)}
\gdef\figin##1{\vskip2in}\gdef\figins##1{\hskip.5in}% blank space instead
\else\message{(FIGURES WILL BE INCLUDED)}%
\gdef\figin##1{##1}\gdef\figins##1{##1}\fi}
\def\DefWarn#1{}
\def\figinsert{\goodbreak\midinsert}
\def\ifig#1#2#3{\DefWarn#1\xdef#1{fig.~\the\figno}
\writedef{#1\leftbracket fig.\noexpand~\the\figno}%
\figinsert\figin{\centerline{#3}}\medskip\centerline{\vbox{
\baselineskip12pt\advance\hsize by -1truein
\noindent\footnotefont{\bf Fig.~\the\figno:} #2}}
%\bigskip
\endinsert\global\advance\figno by1}
%%%
\else
\def\ifig#1#2#3{\xdef#1{fig.~\the\figno}
\writedef{#1\leftbracket fig.\noexpand~\the\figno}%
%\figinsert\figin{\centerline{#3}}\medskip
%\centerline{\vbox{\baselineskip12pt
%\advance\hsize by -1truein\noindent
%\footnotefont{\bf Fig.~\the\figno:} #2}}
%\bigskip\endinsert
\global\advance\figno by1} \fi

\def\id{{1 \kern-.28em {\rm l}}}

\def\pp{{\bf p}}
\def\Z{{\bf Z}}
\def\V{{\bf V}}
\def\w{{\bf w}}
\def\hk{{\hat k}}

\def\K3{{\bf K3}}
\def\journal#1&#2(#3){\unskip, \sl #1\ \bf #2 \rm(19#3) }
\def\andjournal#1&#2(#3){\sl #1~\bf #2 \rm (19#3) }

\def\bar{\overline}
\def\hat{\widehat}

\def\eg{{\it e.g.}}

\def\tilde{\widetilde}

\def\frac#1#2{{#1\over#2}}

\def\inbar{\,\vrule height1.5ex width.4pt depth0pt}
\def\IC{\relax\hbox{$\inbar\kern-.3em{\rm C}$}}
\def\IR{\relax{\rm I\kern-.18em R}}
\def\IP{\relax{\rm I\kern-.18em P}}
\def\Z{{\bf Z}}

%
%%%%%%%%%%%%%%%%%%%%%%%%%%%%%%%%%%%%
%

%
\catcode`\@=11
\def\slash#1{\mathord{\mathpalette\c@ncel{#1}}}
\overfullrule=0pt
\def\r{{\bf r}}
\def\AA{{\cal A}}
\def\BB{{\cal B}}

\def\GG{{\cal G}}

\def\NN{{\cal N}}
\def\OO{{\cal O}}

\def\ZZ{{\cal Z}}

\def\underrel#1\over#2{\mathrel{\mathop{\kern\z@#1}\limits_{#2}}}

\catcode`\@=12

%%%%%%%%%%%%%%%%%%%%%%%%%%%%%%%%%%%%%%%%%%%%%%%%%%%%%%%%%%%%%%

%

\def\det{{\rm det}}

\def\det{{\rm det}}
\def\exp{{\rm exp}}

%%%%%%%%%%%%%%%%%%%%%%%%%%%%%%%%%%%%%%%%%%%%%%%%%%%%%%%%%%%%%%
% new defs:

\def\p{{\partial}}

%\ParnachevHH
\lref\ParnachevHH{
  A.~Parnachev and A.~Starinets,
  ``The Silence of the little strings,''
JHEP {\bf 0510}, 027 (2005). [hep-th/0506144].
%%CITATION = hep-th/0506144%%
}

%\KutasovUF
\lref\KutasovUF{
  D.~Kutasov,
  ``Introduction to little string theory''
}

%\SeibergZK
\lref\SeibergZK{
  N.~Seiberg,
  ``New theories in six-dimensions and matrix description of M theory on T**5 and T**5 / Z(2),''
Phys.\ Lett.\ B {\bf 408}, 98 (1997). [hep-th/9705221].
%%CITATION = hep-th/9705221%%
}

%\AharonyUB
\lref\AharonyUB{
  O.~Aharony, M.~Berkooz, D.~Kutasov, N.~Seiberg
  ``Linear dilatons, NS five-branes and holography,''
JHEP {\bf 9810}, 004 (1998). [hep-th/9808149].
%%CITATION = hep-th/9808149%%
}

%\ItzhakiDD
\lref\ItzhakiDD{
  N.~Itzhaki, J.~M.~Maldacena, J.~Sonnenschein, S.~Yankielowicz
  ``Supergravity and the large N limit of theories with sixteen supercharges,''
Phys.\ Rev.\ D {\bf 58}, 046004 (1998). [hep-th/9802042].
%%CITATION = hep-th/9802042%%
}

%\BoonstraMP
\lref\BoonstraMP{
  H.~J.~Boonstra, K.~Skenderis, P.~K.~Townsend
  ``The domain wall / QFT correspondence,''
JHEP {\bf 9901}, 003 (1999). [hep-th/9807137].
%%CITATION = hep-th/9807137%%
}

%\AharonyTT
\lref\AharonyTT{
  O.~Aharony and T.~Banks,
  ``Note on the quantum mechanics of M theory,''
JHEP {\bf 9903}, 016 (1999). [hep-th/9812237].
%%CITATION = hep-th/9812237%%
}

%\MinwallaXI
\lref\MinwallaXI{
  S.~Minwalla and N.~Seiberg,
  ``Comments on the IIA (NS)five-brane,''
JHEP {\bf 9906}, 007 (1999). [hep-th/9904142].
%%CITATION = hep-th/9904142%%
}

%\GremmHM
\lref\GremmHM{
  M.~Gremm and A.~Kapustin,
  ``Heterotic little string theories and holography,''
JHEP {\bf 9911}, 018 (1999). [hep-th/9907210].
%%CITATION = hep-th/9907210%%
}

%\AharonyDW
\lref\AharonyDW{
  O.~Aharony and M.~Berkooz,
  ``IR dynamics of D = 2, N=(4,4) gauge theories and DLCQ of 'little string theories',''
JHEP {\bf 9910}, 030 (1999). [hep-th/9909101].
%%CITATION = hep-th/9909101%%
}

%\NarayanDR
\lref\NarayanDR{
  K.~Narayan, M.~Rangamani
  ``Hot little string correlators: A View from supergravity,''
JHEP {\bf 0108}, 054 (2001). [hep-th/0107111].
%%CITATION = hep-th/0107111%%
}

%\DeBoerDD
\lref\DeBoerDD{
  P.~A.~DeBoer, M.~Rozali
  ``Thermal correlators in little string theory,''
Phys.\ Rev.\ D {\bf 67}, 086009 (2003). [hep-th/0301059].
%%CITATION = hep-th/0301059%%
}

%\AharonyXN
\lref\AharonyXN{
  O.~Aharony, A.~Giveon, D.~Kutasov
  ``LSZ in LST,''
Nucl.\ Phys.\ B {\bf 691}, 3 (2004). [hep-th/0404016].
%%CITATION = hep-th/0404016%%
}

%\TeschnerFT
\lref\TeschnerFT{
  J.~Teschner,
  ``On structure constants and fusion rules in the SL(2,C) / SU(2) WZNW model,''
Nucl.\ Phys.\ B {\bf 546}, 390 (1999). [hep-th/9712256].
%%CITATION = hep-th/9712256%%
}

%\TeschnerUG
\lref\TeschnerUG{
  J.~Teschner,
  ``Operator product expansion and factorization in the H+(3) WZNW model,''
Nucl.\ Phys.\ B {\bf 571}, 555 (2000). [hep-th/9906215].
%%CITATION = hep-th/9906215%%
}

%\GiveonNS
\lref\GiveonNS{
  A.~Giveon, D.~Kutasov, N.~Seiberg
  ``Comments on string theory on AdS(3),''
Adv.\ Theor.\ Math.\ Phys.\  {\bf 2}, 733 (1998). [hep-th/9806194].
%%CITATION = hep-th/9806194%%
}

%\deBoerPP
\lref\deBoerPP{
  J.~de Boer, H.~Ooguri, H.~Robins, J.~Tannenhauser
  ``String theory on AdS(3),''
JHEP {\bf 9812}, 026 (1998). [hep-th/9812046].
%%CITATION = hep-th/9812046%%
}

%\KutasovXU
\lref\KutasovXU{
  D.~Kutasov, N.~Seiberg
  ``More comments on string theory on AdS(3),''
JHEP {\bf 9904}, 008 (1999). [hep-th/9903219].
%%CITATION = hep-th/9903219%%
}

%\GiveonPX
\lref\GiveonPX{
  A.~Giveon, D.~Kutasov
  ``Little string theory in a double scaling limit,''
JHEP {\bf 9910}, 034 (1999). [hep-th/9909110].
%%CITATION = hep-th/9909110%%
}

%\GiveonTQ
\lref\GiveonTQ{
  A.~Giveon, D.~Kutasov
  ``Comments on double scaled little string theory,''
JHEP {\bf 0001}, 023 (2000). [hep-th/9911039].
%%CITATION = hep-th/9911039%%
}

%\PolchinskiNH
\lref\PolchinskiNH{
  J.~Polchinski, E.~Silverstein
  ``Large-density field theory, viscosity, and '$2k_F$' singularities from string duals,''
Class.\ Quant.\ Grav.\  {\bf 29}, 194008 (2012). [arXiv:1203.1015
[hep-th]].
%%CITATION = arXiv:1203.1015%%
}

%\McGuiganQP
\lref\McGuiganQP{
  M.~D.~McGuigan, C.~R.~Nappi, S.~A.~Yost
  ``Charged black holes in two-dimensional string theory,''
Nucl.\ Phys.\ B {\bf 375}, 421 (1992). [hep-th/9111038].
%%CITATION = IASSNS-HEP-91-57%%
}

%\JohnsonJW
\lref\JohnsonJW{
  C.~V.~Johnson,
  ``Exact models of extremal dyonic 4-D black hole solutions of heterotic string theory,''
Phys.\ Rev.\ D {\bf 50}, 4032 (1994). [hep-th/9403192].
%%CITATION = hep-th/9403192%%
}

%\GiveonGE
\lref\GiveonGE{
  A.~Giveon, E.~Rabinovici, A.~Sever
  ``Beyond the singularity of the 2-D charged black hole,''
JHEP {\bf 0307}, 055 (2003). [hep-th/0305140].
%%CITATION = hep-th/0305140%%
}

%\GiveonZZ
\lref\GiveonZZ{
  A.~Giveon, A.~Sever
  ``Strings in a 2-d extremal black hole,''
JHEP {\bf 0502}, 065 (2005). [hep-th/0412294].
%%CITATION = hep-th/0412294%%
}

%\GiveonJV
\lref\GiveonJV{
  A.~Giveon, D.~Kutasov
  ``The Charged black hole/string transition,''
JHEP {\bf 0601}, 120 (2006). [hep-th/0510211].
%%CITATION = hep-th/0510211%%
}

%\KarabaliDK
\lref\KarabaliDK{
  D.~Karabali, H.~J.~Schnitzer
  ``BRST Quantization of the Gauged WZW Action and Coset Conformal Field Theories,''
Nucl.\ Phys.\ B {\bf 329}, 649 (1990).
%%CITATION = BRX-TH-267%%
}

%\KazamaQP
\lref\KazamaQP{
  Y.~Kazama, H.~Suzuki
  ``New N=2 Superconformal Field Theories and Superstring Compactification,''
Nucl.\ Phys.\ B {\bf 321}, 232 (1989).
%%CITATION = UT-KOMABA-88-8%%
}

%\KazamaUZ
\lref\KazamaUZ{
  Y.~Kazama, H.~Suzuki
  ``Characterization of N=2 Superconformal Models Generated by Coset Space Method,''
Phys.\ Lett.\ B {\bf 216}, 112 (1989).
%%CITATION = UT-KOMABA-88-12%%
}

%\WittenYR
\lref\WittenYR{
  E.~Witten,
  ``On string theory and black holes,''
Phys.\ Rev.\ D {\bf 44}, 314 (1991).
%%CITATION = IASSNS-HEP-91-12%%
}

%\DijkgraafBA
\lref\DijkgraafBA{
  R.~Dijkgraaf, H.~L.~Verlinde, E.~P.~Verlinde
  ``String propagation in a black hole geometry,''
Nucl.\ Phys.\ B {\bf 371}, 269 (1992).
%%CITATION = PUPT-1252%%
}

%\KovtunEV
\lref\KovtunEV{
  P.~K.~Kovtun, A.~O.~Starinets
  ``Quasinormal modes and holography,''
Phys.\ Rev.\ D {\bf 72}, 086009 (2005). [hep-th/0506184].
%%CITATION = hep-th/0506184%%
}

%\GiulianiV
\lref\GiulianiV{
G.F. Giuliani and G. Vignale, "Quantum theory of the
electron liquid," Cambridge University Press, 2005.
}

%\PolchinskiRQ
\lref\PolchinskiRQ{
  J.~Polchinski,
  ``String theory. Vol. 1: An introduction to the bosonic string,''
Cambridge, UK: Univ. Pr. (1998) 402 p. }

%\PolchinskiRR
\lref\PolchinskiRR{
  J.~Polchinski,
  ``String theory. Vol. 2: Superstring theory and beyond,''
Cambridge, UK: Univ. Pr. (1998) 531 p. }

%\WittenMM
\lref\WittenMM{
  E.~Witten,
  ``On Holomorphic factorization of WZW and coset models,''
Commun.\ Math.\ Phys.\  {\bf 144}, 189 (1992).
%%CITATION = IASSNS-HEP-91-25%%
}

%\CallanIA
\lref\CallanIA{
  C.~G.~Callan, Jr., E.~J.~Martinec, M.~J.~Perry and D.~Friedan,
  ``Strings in Background Fields,''
Nucl.\ Phys.\ B {\bf 262}, 593 (1985).
%%CITATION = PRINT-85-0734 (PRINCETON)%%
}

%\JohnsonKV
\lref\JohnsonKV{
  C.~V.~Johnson,
  ``Heterotic Coset Models,''
Mod.\ Phys.\ Lett.\ A {\bf 10}, 549 (1995). [hep-th/9409062].
%%CITATION = hep-th/9409062%%
}

%\KaminskiDH
\lref\KaminskiDH{
  M.~Kaminski, K.~Landsteiner, J.~Mas, J.~P.~Shock and J.~Tarrio,
  ``Holographic Operator Mixing and Quasinormal Modes on the Brane,''
JHEP {\bf 1002}, 021 (2010). [arXiv:0911.3610 [hep-th]].
%%CITATION = arXiv:0911.3610%%
}

%\GoykhmanVY
\lref\GoykhmanVY{
  M.~Goykhman, A.~Parnachev and J.~Zaanen,
  ``Fluctuations in finite density holographic quantum liquids,''
JHEP {\bf 1210}, 045 (2012). [arXiv:1204.6232 [hep-th]].
%%CITATION = arXiv:1204.6232%%
}

%\DasEI
\lref\DasEI{
  S.~R.~Das and S.~P.~Trivedi,
  ``Three-brane action and the correspondence between N=4 Yang-Mills theory and anti-De Sitter space,''
Phys.\ Lett.\ B {\bf 445}, 142 (1998). [hep-th/9804149].
%%CITATION = hep-th/9804149%%
}

%\FerraraBP
\lref\FerraraBP{
  S.~Ferrara, M.~A.~Lledo and A.~Zaffaroni,
  ``Born-Infeld corrections to D3-brane action in AdS(5) x S(5) and N=4, d = 4 primary superfields,''
Phys.\ Rev.\ D {\bf 58}, 105029 (1998). [hep-th/9805082].
%%CITATION = hep-th/9805082%%
}

%%%%%%%%%%%%%%%%%%%%%%%%%%%%%%%%%%%%%%%%%%%%%%%%%%%
%%%%%%%%%%%%%%%%%%%%%%%%%%%%%%%%%%%%%%%%%%%%%%%%%%%
\Title{} {\vbox{\centerline{Stringy holography at finite density}
%\bigskip
%\centerline{}
}}
\bigskip

\centerline{\it Mikhail Goykhman  and Andrei Parnachev}
\bigskip
\smallskip
\centerline{Institute Lorentz for Theoretical Physics, Leiden
University} \centerline{P.O. Box 9506, Leiden 2300RA, The
Netherlands}
\smallskip

\vglue .3cm

\bigskip

\let\includefigures=\iftrue

\noindent We consider an exactly solvable worldsheet string theory
in the background of a black brane with a gauge field flux.
Holographically, such a system can be interpreted as a field theory
with finite number of degrees of freedom at finite temperature and
density. This is to be contrasted with more conventional holographic
models which involve gravity in the bulk  and possess infinite
number of degrees of freedom and mean field critical exponents. We
construct closed string vertex operators which holographically
represent the $U(1)$ gauge field and the stress energy tensor and
compute their two-point functions. At finite temperature and
vanishing charge density the low energy excitations are  described by
hydrodynamics. As the density is raised, the system behaves like a
sum of two noninteracting fluids. We find low-energy excitations in
the shear and sound channels of each fluid.

\bigskip

\Date{April 2013}

\newsec{Introduction}

In the usual AdS/CFT setting gauge theory on the boundary has a dual
description in terms of closed string theory in the bulk. Most
often, a limit of small curvature is taken to yield a low energy theory
of strings, supergravity. In the $\NN=4$ supersymmetric Yang-Mills
case this limit implies strong 't Hooft coupling of field theory.
%In some bulk backgrounds the string theory worldsheet computations
%are actually simpler than gravity computations.
A distinct example of  non-gravitational theory with a holographically dual
description  is the Little String Theory
\refs{\SeibergZK,\AharonyUB}. It can be viewed as the theory of $N$
coincident NS5-branes, taken at vanishing string coupling, $g_s=0$,
where the bulk degrees of freedom decouple. The coupling constant of
the low-energy $U(N)$ gauge degrees of freedom, living on the
NS5-branes world-volume, stays unaffected by taking this limit, and
is equal to $g_5=\ell_s$, where $\ell_s$ is string length in
type-IIB string theory (see \KutasovUF\ for a review).

The holographic dual of the Little String Theory
\refs{\AharonyUB,\ItzhakiDD,\BoonstraMP} is the theory of closed
strings in the background of NS5-branes, with the geometry
$R^{5,1}\times R^\phi\times SU(2)_N$, the two-form field and the
linear dilaton.
%$\Phi=-\frac{Q}{2}\phi$, where
%$Q=\frac{2}{\sqrt{N}\ell_s}$ and $\ell_s$ is the string length.
The CFT on $SU(2)$ is described by WZW action at level $N$. The bulk
physics (in the double scaling limit) can be reformulated as the
string theory on $R^{5}\times \frac{SL(2,R)_N}{U(1)}\times SU(2)_N$
space-time. This is due to the fact that the gauged WZW model on
$SL(2,R)_N/U(1)$ gives rise to the classical ``cigar" geometry of
the two-dimensional black hole with the asymptotically linear dilaton
\refs{\WittenYR-\DijkgraafBA}.
In the large $N$ limit the bulk theory reduces to supergravity\foot{The radius of
the $SU(2)$ sphere is $R_{sph}=\sqrt{N}\ell_s$. Therefore the large
$N$ limit is equivalent to the limit of small $\ell_s/R_{sph}$.}.

Generally one expects that a lot of nontrivial physics drastically
simplifies in the limit of infinitely many degrees of freedom (large $N$ limit), both in the boundary field theory
and from the dual bulk perspective. For example, one expects the
large $N$ physics of a field theory at finite temperature and
density to have ``classical" nature, resulting, in particular, in
the mean field critical exponents. (Another recent example of this
is given by the stringy nature of finite-momentum and zero-frequency
singularity of the current-current two-point functions, observed in
\PolchinskiNH, where the results of
\refs{\TeschnerFT\TeschnerUG\GiveonNS\deBoerPP\KutasovXU\GiveonPX-\GiveonTQ}
were extensively used.)
%The large N limit can
%make certain effects disappear from the field theory.
%Correspondingly, these effects become invisible in the bulk theory
%as well.
%Let us consider an example of an essentially finite N effect. One of
%the exactly known results
%\refs{\TeschnerFT\TeschnerUG\GiveonNS\deBoerPP\KutasovXU\GiveonPX-\GiveonTQ}
%of classical string theory concerns the two-point function of the
%primary fields $V_{j,m,\bar m}$. These fields are the vertex
%operators of the ground state of string theory with $SL(2,R)$ target
%space-time. As it was observed in \PolchinskiNH\ this two-point
%function, when applied in the context of holographic correspondence,
%has a pole at finite momentum and zero frequency. This is
%essentially finite N result, and therefore is of the stringy nature.

The low energy excitations in Little String Theory at finite
temperature have been considered in \ParnachevHH\ \foot{See also
\eg\
\refs{\AharonyTT\MinwallaXI\MinwallaXI\AharonyDW\NarayanDR\DeBoerDD-\AharonyXN}
for some preceding holographic study of the Little String Theory.}.
The closed string description involves  the gauged WZW (gWZW) action
with the $SL(2,R)/U(1)$ target space-time and ${\cal N}=2$
worldsheet supersymmetry. In \ParnachevHH\ the two-point functions
of the stress-energy tensor and the $U(1)$ current have been
computed holographically; their pole structure indicates the
presence of hydrodynamic  modes. This has been also verified by
solving fluctuation equations in supergravity approximation in the
background of a large number of NS5-branes.

In this paper we study string theory in the background
of a direct product of the two-dimensional charged black hole
\McGuiganQP\ and flat space. The string theory in the
two-dimensional charged black hole background is described by the
gWZW action with the $\frac{SL(2,R)\times U(1)_x}{U(1)}$ target
space-time \refs{\JohnsonJW\GiveonGE-\GiveonZZ}. Here $U(1)_x$ is a
compact circle, which is Kaluza-Klein reduced, and $U(1)$ subgroup
of $SL(2,R)\times U(1)_x$ is gauged asymetrically. The left-moving
sector of the gauged $U(1)$ is a linear combination of the
left-moving sector of the $U(1)_x$ and the left-moving sector of the
$U(1)$ subgroup of the $SL(2,R)$. The coefficient of this linear
combination determines the charge to mass ratio of the resulting
black hole. The right-moving sector of the gauged $U(1)_x$ is the
right-moving sector of the $U(1)$ subgroup of the $SL(2,R)$.

This bulk system is holographically dual to the boundary quantum
field theory at finite temperature and charge density. (One can
think of the resulting system as  little string theory at finite
density, but we do not study the field theoretic interpretation here
in detail.) The inverse temperature is equal to
$\beta=2\pi\sqrt{\hk}\frac{\cos^2(\psi/2)}{\cos\psi}$, where
$\psi\in [0,\pi/2]$ is the parameter of asymmetric gauging. The
finite charge density in the field theory is described
holographically by the background $U(1)$ potential in the bulk,
$A_t(u)\simeq -\frac{q}{u}$, where $q=M\sin\psi$ is the charge and
$M$ is the mass of the black hole; $u$ is the radial coordinate in
the bulk.

The vertex operator of the string ground state in this model was
constructed in \GiveonGE. In this paper we construct the vertex
operators which describe massless closed string excitations in this
model, which constitute the NS-NS sector of type-II supergravity. We
also construct the gauge field vertex operators, which are obtained
by Kaluza-Klein reduction on $U(1)_x$ from graviton and
antisymmetric tensor field vertex operators. The graviton in the
bulk is dual to the stress-energy tensor on the boundary; the gauge
field in the bulk is dual to the charge current on the boundary.
%We find it useful that instead of considering graviton $G_{\mu\nu}$,
%anti-symmetric tensor $B_{\mu\nu}$ and gauge fields $A_\mu =G_{\mu
%x}$ and $B_\mu =B_{\mu x}$ one considers the vertex operators
%$R_{\mu\nu}=G_{\mu\nu}-B_{\mu\nu}$, $U_\mu=G_{\mu x}-B_{\mu x}$,
%which we call R-system, and the vertex operators
%$S_{\mu\nu}=G_{\mu\nu}+B_{\mu\nu}$, $W_\mu=G_{\mu x}+B_{\mu x}$,
%which we call S-system. The R-system is decoupled from the S-system.
We study the low energy excitations of the system by computing
holographically the two-point functions for the charge current and
the stress-energy tensor and reading off the dispersion relation
from their poles.
%We find that both R-system and S-system
%generally exhibit hydrodynamic and diffusion poles. The poles of the
%R-system do not depend on the charge to mass ratio. At zero
%temperature (for the extremal black hole), the poles of the S-system
%disappear, and at zero charge the poles of the S-system coincide
%with the poles of the R-system. We confirm these results by studying
%corresponding fluctuation equations in type-II gravity.
We find two distinct gapless modes in the shear channel;
the dispersion relation of one of them  is independent of the
charge to mass ratio of the black hole. The two modes
merge in the limit of vanishing charge, producing the
shear mode which was observed in  \ParnachevHH.
We confirm these results by solving
fluctuation equations of the type-II supergravity.
The situation in the sound channel is similar.

Finally we study fluctuation equations in the low-energy limit in
heterotic gravity \McGuiganQP. We find one gapless mode in the shear
channel. Comparing this result with the thermodynamics of the
charged black hole \GiveonJV\ we find that the ratio of shear
viscosity to entropy density is equal to $\eta/s=1/(4\pi)$,
independently of the charge to mass ratio of the black hole.

The rest of the paper is organized as follows. In section 2 we
review the thermodynamics of the two-dimensional charged black hole
and derive the dispersion relation of the shear hydrodynamic mode.
In section 3 we apply the BRST quantization method of the coset
models, and the covariant quantization of the string to construct
the holomorphic and anti-holomorphic physical vertex operators of
the massless states on the $\frac{SL(2,R)\times U(1)}{U(1)}$ coset.
In section 4 we use these vertex operators and write down the vertex
operators of graviton, antisymmetric tensor field and gauge fields.
In that section we also compute the two-point functions of these
vertex operators and discuss the low-energy excitation modes. We
also briefly discuss finite-momentum and zero-frequency singularity
of the correlation functions. In section 5 we solve fluctuation
equations in type-II supergravity to verify the dispersion
relations, derived in section 4. We discuss our results in section 6.
Appendix A is devoted to a review
of some rudimentary conformal field theory and derivation of the
gWZW action on the $\frac{SL(2,R)\times U(1)}{U(1)}$ coset. In
Appendix B we solve fluctuation equations in heterotic gravity. We
find one mode in the shear channel. Matching its dispersion relation
to the one, written in section 2, we obtain that $\eta/s=1/(4\pi)$
for any charge to mass ratio.

\newsec{Thermodynamics of the charged black hole}

The metric of the two-dimensional charged black hole \McGuiganQP\
with mass $M$ and charge $q$ in suitable coordinates can be written
as \refs{\GiveonZZ,\GiveonJV}
\eqn\cbhmetr{ds^2=-f(u)dt^2+\frac{\hat
k}{4}\frac{du^2}{u^2f(u)}\,,\quad
f(u)=\frac{(u-u_+)(u-u_-)}{u^2}\,,\quad u_\pm =M\pm\sqrt{M^2-q^2}}
with the background $U(1)$ gauge field and the dilaton field being
equal to
\eqn\cbhbfl{ A_t(u)=q\left(\frac{1}{u_+}-\frac{1}{u}\right)\,, \quad
\Phi=\Phi_0-\frac{1}{2}\log\left(\frac{u\sqrt{\hk}}{2}\right)\,,\quad
\Phi_0=-\frac{1}{2}\log\left(\frac{Mu_+\sqrt{\hk}}{u_++u_-}\right)\,.}
The gauge potential vanishes at the outer horizon, $A_t(u_+)=0$.
Define parameter $\psi$ by the equation
\eqn\rmrp{\frac{u_-}{u_+}=\tan^2\frac{\psi}{2}}
%

%Let us use grand canonical ensemble description of the charged black
%hole.
For the full description of thermodynamics of two-dimensional
charged black hole the reader is referred to \GiveonJV, we just
review their results which are useful for us.
%For given values of the inverse temperature $\beta$ and the
%chemical potential $\mu$ we consider ensemble of the charged black
%hole states with various values of energy (mass) and charge. The
%grand canonical partition sum is equal to the trace of the Boltzmann
%factor over the ensemble of states
%%
%\eqn\partZ{Z={\rm Tr}\, e^{-\beta (H-\mu \QQ)}\,,}
%%
%where each eigenvalue of $H$ gives mass and each eigenvalue of $\QQ$
%gives charge of the state in the ensemble.
It is convenient for further purposes to denote the background
dilaton slope (see eq. \cbhbfl) as $Q=2/\sqrt{\hat k}$. Requiring
the metric \cbhmetr\ near external horizon $u=u_+$ to be regular, we
find the temperature of the charged black hole
\eqn\betab{\beta=\frac{4\pi}{Q}\frac{u_+}{u_+-u_-}
=\frac{4\pi}{Q}\frac{\cos^2\frac{\psi}{2}}{\cos\psi} \,.}
Asymptotical $u\gg 1$ value of the gauge potential (see eq. \cbhbfl)
is equal to the chemical potential:
\eqn\chpot{\mu
=\frac{q}{u_+}=\sqrt{\frac{u_-}{u_+}}=\tan\frac{\psi}{2}}
%
%Instead of specifying the value of temperature of the grand
%canonical ensemble we therefore can specify the value of dilaton
%slope $Q$.
The entropy of the two-dimensional charged black hole is given by
\GiveonJV\
\eqn\Scbh{S_{bh}(M,q)=\frac{2\pi}{Q}(M+\sqrt{M^2-q^2})}
%
%Each state in the partition sum \partZ\ is characterized by its mass
%$M$ and charge $q$. Instead of the charge $q$, due to $q=M\sin\psi$,
%one can characterize each state by the value of~$\psi$. Then for the
%large entropy, as it was observed in \GiveonJV, the states with
%$\psi$ satisfying \rmrp\ contribute the most to the partition sum
%\partZ.
Using \Scbh\ and evaluating the grand canonical partition sum $\ZZ$
one observes \GiveonJV\ that the grand canonical potential
$\Omega\sim -\log \ZZ$ vanishes, and therefore the pressure
vanishes.

Consider black brane background space-time $CBH_2\times R^{d-1}$,
which is a direct product of the two-dimensional charged black hole
and flat $d-1$-dimensional space. Denote by $X$ the direction of
$R^{d-1}$ of propagation of all the excitation, and denote by $Y$
some transverse direction of $R^{d-1}$. In the shear channel
excitation modes appear as poles of the two-point function $\langle
T_{XY}T_{XY}\rangle$ of the stress-energy tensor $T_{MN}$, with the
dispersion relation of the low-energy mode given by
\eqn\shzeroPro{\omega =-\frac{i\eta}{(M+P)/V}p^2}
where $p$ is the momentum and $\omega$ is the frequency of the mode;
$\eta$ is the shear viscosity, $M/V$ and $P/V$ are energy and
pressure densities.

Because for the two-dimensional charged black hole the pressure
vanishes, we obtain
\eqn\shzeroPr{\omega =-\frac{i\eta}{M/V}p^2}
Using \Scbh\ one can express,
\eqn\Mentrcon{M=\frac{QS_{bh}}{2\pi (1+\cos\psi)}}
and therefore the hydrodynamics predicts a shear pole with the
dispersion relation
\eqn\shzerPt{\omega =-4\pi i\frac{\eta}{s}\frac{\sqrt{\hat
k}\cos^2(\psi/2)}{2}p^2\,,}
where $s=S/V$ is the entropy density. Bellow we are going to compare
this result with the computation in heterotic gravity and derive the
value of $\eta/s$.

%Using $q=M\sin\psi$ one can replace in the partition function
%\partZ\ the sum over $q$ by the sum over $\psi$:
%%
%\eqn\Zintm{Z\simeq\int d\psi\int dM e^{S_{bh}(M,M\sin\psi)-\beta M
%(1-P\sin\psi)}}
%%
%Using \Scbh\ one can express,
%%
%\eqn\Mentrcon{M=\frac{QS_{bh}}{(2\pi (1+\cos\psi)}}
%%
%and therefore
%%
%\eqn\ZSm{Z\simeq\int d\psi\int dM
%\exp\left(\left(1-\frac{Q\beta}{2\pi}\frac{1-P\sin\psi}{1+\cos\psi}
%\right)S_{bh}(M,M\sin\psi)\right)}
%%
%Consequently for large entropy dominant contribution to the
%partition sum comes from the terms with charge $q$, and
%corresponding value of $\psi$, which minimizes the function
%$\frac{1-P\sin\psi}{1+\cos\psi}$. Such states are the states with
%$P=\tan\frac{\psi}{2}$. We can neglect contributions from other
%$\psi$, therefore integrating \ZSm\ we obtain
%%
%\eqn\ZSmGrpot{Z\simeq
%\exp\left(\left(1-\frac{Q\beta}{2\pi}\frac{1-P\sin\psi}{1+\cos\psi}
%\right)S_{bh}(M,M\sin\psi)\right)_{\tan\frac{\psi}{2}=P}}
%%
%Plugging expression \betab\ for the temperature, one obtains, that
%grand canonical potential $\Omega\sim -\log Z$ vanishes.

%This means that pressure vanishes, and consequently shear viscosity
%dispersion relation acquires the form
%%
%\eqn\shzeroPr{\omega =-\frac{i\eta}{M/V}p^2}
%%
%Therefore due to \Mentrcon\
%%
%\eqn\shzerPt{\omega =-4\pi
%i\frac{\eta}{s}\frac{\cos^2(\psi/2)}{Q}p^2\quad\Rightarrow\quad
%\omega =-4\pi i\frac{\eta}{s}\frac{\sqrt{\hat
%k}\cos^2(\psi/2)}{2}p^2\,,}
%%
%where $s=S/V$ is the entropy density.

\newsec{The physical state conditions and vertex operators}

\subsec{BRST quantization}

%String theory with the $\frac{SL(2,R)\times U(1)}{U(1)}$ coset
%target space-time is described by the gWZW action.
The gauging of the $U(1)$ subgroup from the $SL(2,R)\times U(1)$
group in the gWZW model on the $\frac{SL(2,R)\times U(1)}{U(1)}$
coset is realized by adding the $U(1)$ non-dynamical gauge field to
the system, and adding corresponding action terms to the
$SL(2,R)\times U(1)$ WZW action. The $U(1)$ subgroup is gauged
left-right asymetrically, and anomaly-free condition must be
satisfied. The details of the construction are reviewed in Appendix
A. The end product is the gWZW action
\eqn\gWZWkk{S_g=S[g]+\frac{1}{2\pi}\int d^2z\p x\bar \p
x+\frac{1}{2\pi}\int d^2z \left[A\,\tilde k+\bar A\, k    +A\bar
A\left(2+{\rm Tr}(g^{-1}\sigma ^3g\sigma^3)\cos\psi\right)\right]}
Here we have denoted the currents of the gauged $U(1)$ subgroup as
\eqn\kdef{k=\sqrt{\hat k}{\rm Tr}(\p gg^{-1}\sigma
^3)\cos\psi+2\sin\psi\p x=\frac{2}{\sqrt{\hat
k}}j^3\cos\psi+2\sin\psi\p x}
\eqn\barkdef{\tilde k=\sqrt{\hat k}{\rm Tr}\, (g^{-1}\bar\p g\sigma
^3)=-\frac{2}{\sqrt{\hat k}}\tilde j^3\,.}

To determine physical spectrum of the quantum model on the
$\frac{SL(2,R)\times U(1)}{U(1)}$ coset, we are going to use BRST
quantization method \KarabaliDK\ (see \DijkgraafBA\ where this
method was applied to build the $SL(2,R)/U(1)$ model). The path
integral for the theory is
\eqn\pathi{Z=\int [dG][dA][d\bar A]\exp\left(-S_{g}[G,A,\bar
A]\right)}
\eqn\pathit{=\int [dG][d u][d
v]\det\,\p\,\det\,\bar\p\;\exp\left(-S[G]+S[w]\right)}
Represent functional determinants in terms of the gauge ghost fields
\eqn\fdet{\det\,\p\,\det\,\bar\p=\int [db][dc][d\tilde b][d\tilde
c]\exp\left(-\frac{1}{2\pi}\int d^2z(b\bar\p c+\tilde b\p\tilde
c)\right)}
Ghosts satisfy OPEs
\eqn\ghosts{c(z)b(w)\sim\frac{1}{z-w}+\cdots\,,\quad\quad\tilde
c(\bar z)\tilde b(\bar w)\sim\frac{1}{\bar z-\bar w}+\cdots}
Fix the gauge symmetry, for concreteness fix $v=1$, therefore $u=w$.
Consequently the path integral is given by
\eqn\pathi{Z=\int [dG][dw][db][dc][d\tilde b][d\tilde
c]\exp\left(-S[G]+S[w]-\frac{1}{2\pi}\int d^2z(b\bar\p c+\tilde b\p
\tilde c)\right)}
and the total action is given by
\eqn\Sq{S_q=S[G]-S[w]+\frac{1}{2\pi}\int d^2z(b\bar\p c+\tilde b\p
\tilde c)}
Notice from this action that the correlation function for $w$ has
the wrong sign:
\eqn\wwcor{\langle\p w(z_1)\p w(z_2)\rangle =\frac{1}{2(z_1-z_2)^2}}

Perform variations $(\delta G,\delta w,\delta b)$ in the action \Sq,
\eqn\Sqvar{\delta S_q=-\frac{1}{2\pi}\int d^2z\left[\hat k{\rm
Tr}(\p GG^{-1}\bar\p (G^{-1}\delta G))+2\p w\bar \p\delta w -\delta
b\bar\p c\right]}
For the transformations with a local Grassmann parameter $\eta$:
%\foot{Due to $\langle b(z)b(0)\rangle =0$, the field $b$ can only be
%transformed by a null operator $k+2\p w$, see bellow.}
%
\eqn\brstsq{\delta G=\eta cGT_L\,,\quad \delta w=\eta c\,,\quad
\delta b=\eta (k+2\p w)}
we therefore obtain
\eqn\deltase{\delta S_q=\frac{1}{2\pi}\int d^2 z(\bar\p \eta)\,
c(k+2\p w)}
If $\eta$ is a global parameter, then $S_q$ is invariant.
Corresponding transformation is BRST symmetry transformations. Then
from \deltase\ for anti-holomorphic $\eta(\bar z)$ we can read off
holomorphic component of the corresponding conserved Noether
current:
\eqn\jbrst{j_{BRST}=c(k+2\p w)}
Notice that
\eqn\bbcor{\langle (k(z_1)+2\p w(z_1))(k(z_2)+2\p w(z_2))\rangle
=0\,.}
Therefore corresponding BRST charge
\eqn\qbrst{Q_{BRST}=\frac{1}{2\pi i}\oint dz j_{BRST}}
is nilpotent. Similarly one finds the anti-holomorphic component of
the BRST current
\eqn\jbrstc{\tilde j_{BRST}=\tilde c(\tilde k+2\bar\p\tilde w)}

Physical states of the $\frac{SL(2,R)\times U(1)}{U(1)}$ coset model
are the BRST-closed states of the $SL(2,R)\times U(1)$ model:
\eqn\physcond{Q_{BRST}|{\rm phys}\rangle =0\,,\quad \tilde
Q_{BRST}|{\rm phys}\rangle =0\,,}
and are defined up to BRST-exact states.

Denote null bosonic currents as
\eqn\jdef{J=k+2\p w\,,\quad \quad\tilde J=\tilde k+2\bar\p\tilde w}
BRST physical state conditions \physcond\ therefore become
\eqn\physcondj{J_n|{\rm phys}\rangle =0\,,\;\; n\geq 0\,,\quad\tilde
J_n|{\rm phys}\rangle =0\,,\;\; n\geq 0}
The BRST-exact massless state is obtained by acting with $J_{-1}$
and $\tilde J_{-1}$ on the BRST-closed ground state.

\subsec{Ground state vertex operator}

The ground state vertex operator $V_t$ of the $\frac{SL(2,R)\times
U(1)}{U(1)}$ model was constructed in \GiveonGE\ as a ground state
vertex operator of the $SL(2,R)\times U(1)$ model invariant under
gauge $U(1)$ transformations. This vertex operator describes a
tachyon, which due to GSO projection is projected out of the NS-NS
sector. The lowest NS states are massless, and they are described by
the vertex operator $V^M=j^M_{-1}V_t$ (and similarly for
anti-holomorphic vertex operator). We derive these massless vertex
operators bellow in this section. In this subsection we find it
useful to reproduce the result of \GiveonGE\ using BRST
quantization, developed in the previous subsection.

Suppose $\Phi (z,\bar z)$ is a vertex operator on $SL(2,R)\times
U(1)$. Then
\eqn\vphi{V_t(z,\bar z)=\Phi (z,\bar z)
\exp\left(im_Lw_L+im_Rw_R\right)}
where
\eqn\wlr{w(z,\bar z)=w_L(z)+w_R(\bar z)}
is a vertex operator on the coset $\frac{SL(2,R)\times U(1)}{U(1)}$
if the physical state condition \physcond\ are satisfied. We obtain
\eqn\pscc{k_0\cdot V_t(z,\bar z)=-im_L V_t(z,\bar z)}
\eqn\pscct{\tilde k_0\cdot V_t(z,\bar z)=-im_R V_t(z,\bar z)}
The non-compact $w$-circle contains only momentum modes and does not
contain any winding modes, therefore
\eqn\windcond{m_L=\frac{M}{R}-WR\,,\quad
m_R=\frac{M}{R}+WR\quad\Rightarrow\quad m_L=m_R}
Let us denote $m_L=m_R=N$.  The ground state on $SL(2,R)\times U(1)$
is described by the vertex operator \foot{Define $x\sim x+\pi$, so
that $n_{L,R}$ are integers. The $V_{jm\bar m}$ is the $SL(2,R)$
ground state primary field, see details in Appendix A.}
\eqn\tgstate{\Phi (z,\bar z)=V_{jm\bar m}e^{2in_Lx_L+2in_Rx_R}\,,}
therefore
\eqn\pscca{k_0\cdot \Phi(z,\bar
z)=2\left(\frac{m\cos\psi}{\sqrt{\hat
k}}-in_L\sin\psi\right)\Phi(z,\bar z)}
\eqn\psccaa{\tilde k_0\cdot \Phi(z,\bar z)=-\frac{2\bar
m}{\sqrt{\hat k}}\Phi(z,\bar z)}
The BRST physical state conditions \pscc, \pscct\ due to \windcond\
therefore imply
\eqn\mN{2\left(\frac{m\cos\psi}{\sqrt{\hat
k}}-in_L\sin\psi\right)=-iN}
\eqn\barmN{\frac{2\bar m}{\sqrt{\hat k}}=iN}
and consequently
\eqn\pscg{m\cos\psi+\bar m-i\sqrt{\hat k}n_L\sin\psi=0\,,}
as was derived in \GiveonGE.

\subsec{Vertex operators of massless states in type-II superstring
theory}

The stress-energy tensor, which follows from the action \Sq, has the
following holomorphic (left-moving) component \foot{The term with
$w$ corresponds to the coset Kazama-Suzuki construction
\refs{\KazamaQP,\KazamaUZ}, where $T_{G/H}=T_G-T_H$, in the
following way. From BRST condition due to \jbrst\ one obtains,
schematically, $\p w=-\frac{1}{2}k$. Therefore contribution of $w$
to the stress-energy tensor $T(z)$ is $T_w(z)=\p w\p w
=\frac{1}{4}kk$. Then we expect $T_H=-T_w$, which is indeed the
case: $T_H(z)k(0)=k(z)/z^2$.}
\eqn\Tz{T(z)=\frac{1}{\hat k}\eta_{AB}j^Aj^B-\p x\p x+\p w\p w}
and similarly for anti-holomorphic component. Therefore
$\OO((z-w)^{-2})$ terms of the OPEs of the stress-energy tensor and
the ground state primary $V_t$ are given by
\eqn\Tzvtope{T(z)V_t(w,\bar
w)=\frac{1}{(z-w)^2}\left(-\frac{j(j+1)}{\hat
k}+\frac{n_L^2-m_L^2}{2}\right)V_t(w,\bar w)+\cdots}
\eqn\Tzvtopet{\tilde T(\bar z)V_t(w,\bar w)=\frac{1}{(\bar z-\bar
w)^2}\left(-\frac{j(j+1)}{\hat
k}+\frac{n_R^2-m_R^2}{2}\right)V_t(w,\bar w)+\cdots}
In what follows we are going to perform Kaluza-Klein reduction of
the $U(1)_x$ circle, therefore $n_L=n_R=0$.

In this paper we are interested in the NS-NS vertex operators of the
massless closed string excitations in type-II superstring theory.
These operators in the $(-1,-1)$ picture are constructed as
(anti)symmetrized direct products of massless holomorphic $\V^\mu
(z)=e^{-\varphi} \psi ^\mu _{-1/2}\cdot V_t$ and anti-holomorphic
$\tilde \V^\mu (\bar z) =e^{-\tilde\varphi}\tilde\psi ^\mu
_{-1/2}\cdot V_t$ vertex operators. Here $\psi ^\mu (z)$ and $\tilde
\psi^\mu (\bar z)$ are worldsheet fermions, and
$\varphi\,,\tilde\varphi$ are bosonized superconformal ghosts. The
only non-trivial super-Virasoro physical state condition, which one
needs to impose on the massless states, is $G_{1/2}\cdot \V^\mu
(z)=0$, and similarly for anti-holomorphic vertex operator, where
$G(z)=\sum _r\, G_r/z^{r+3/2}$ is the supercurrent.

The other option, which is what we are going to use in this paper,
is to consider vertex operators $V^\mu$ and $\tilde V^\mu$ in
zero-ghost picture. They are obtained from $(-1,-1)$ picture vertex
operators $\V^\mu$ and $\tilde\V^\mu$ by acting with the picture
changing operators $e^\varphi G$ and $e^{\tilde\varphi}\tilde G$. As
a result one obtains vertex operator in zero-ghost picture
\eqn\picch{V^\mu =G_{-1/2}\cdot\psi_{-1/2}\cdot
V_t=(j^\mu_{-1}+p\cdot\psi_{-1/2}\psi^\mu _{-1/2})\cdot V_t\,,}
where $j^\mu$ is the current, supersymmetric to the fermion
$\psi^\mu$, and $p^\mu$ is the momentum of the state. Similar
expression is true for anti-holomorphic vertex operator. The only
non-trivial super-Virasoro constraint which one should impose in
zero-ghost picture is $L_1\cdot V^\mu =0$. Moreover, the $L_1$ here
is actually the amplitude of the stress-energy tensor $L_1^{(b)}$
for only bosonic modes: in the r.h.s. of \picch\ contribution of
fermions is automatically annihilated by the fermionic stress-energy
tensor amplitude $L_1^{(f)}=\psi^\nu _{1/2}\psi_{\nu 1/2}$.
Therefore instead of studying massless NS-NS states in type-II
superstring theory we can study gravity multiplet in bosonic string
theory.

\subsec{(Anti)holomorphic vertex operators of massless modes in the
$R\times\frac{SL(2,R)}{U(1)}$ coset model}

In this subsection we review the construction of (anti)holomorphic
vertex operators~\ParnachevHH, describing massless
(right-)left-moving excitations in the gWZW model on
$R\times\frac{SL(2,R)}{U(1)}$ \refs{\WittenYR-\DijkgraafBA}. The
classical background is the two-dimensional black hole with the
linear dilaton in a direct product with a real line. The real line
is parametrized by the flat coordinate $X$, which we choose as a
direction of propagation of all the excitations. The momentum is
equal to $p$.

The authors of~\ParnachevHH\ considered graviton vertex operator in
$(-1,-1)$ picture on the worldsheet with  $\NN=2$ supersymmetry. We
perform a picture changing and consider vertex operators in
zero-ghost picture. Due to noted in the previous subsection, we can
actually study bosonic string and then make contact with the results
of \ParnachevHH.

Without loss of generality let us focus on holomorphic vertex
operators. The ground state vertex operator of the
$R\times\frac{SL(2,R)}{U(1)}$ coset theory is
\eqn\Rtvo{V_t=e^{ipX}e^{iNw}V_{jm}}
This state must be closed under the action of the null $U(1)$ BRST
current,
\eqn\Ruonbrst{J=j^3-\sqrt{\hk}\p w\,,}
which imposes the condition
\eqn\Rtpsc{iN=\frac{2m}{\sqrt{\hk}}}

The most general holomorphic vertex operator of the massless state
(which is a gauge field from the space-time point of view) on
$R\times\frac{SL(2,R)}{U(1)}$ is
\eqn\Rgengf{V^\chi=(a_+j^+_{-1}j^-_0+a_-j^-_{-1}j^+_0+a_3j^3+a_w\p
w+a_X\p X)V_t}
Mass-shell Virasoro constraint (see \Tzvtope) gives
\eqn\RmsVc{L_0V^\chi=V^\chi\quad\Rightarrow\quad -\frac{j(j+1)}{\hk
-2}+\frac{p^2-N^2}{4}=0}
Closeness of \Rgengf\ w.r.t. $J_1$ (see \Ruonbrst) reduces the
number of parameters by one, giving the most general BRST-closed
state
\eqn\Rgengfgencl{V^\chi{=}(a_+j^+_{-1}j^-_0{+}a_-j^-_{-1}j^+_0{+}\frac{2}{\sqrt{\hk}}(a_+(m{+}j)(m{-}1{-}j)
{-}a_-(m{-}j)(m{+}1{+}j))\p w{+}AJ{+}a_X\p X)V_t}
Also $V^\chi$ is defined up to BRST-exact state $JV_t$, which makes
one more parameter unphysical, leaving us with a gauge field in
three dimensional $R\times\frac{SL(2,R)}{U(1)}$ with three
polarization parameters.

Gauge field in three dimensions has one transverse physical d.o.f.
Two of the three d.o.f. are eliminated in the following way. First,
we impose Virasoro constraint $L_1V^\chi=0$. Second, the state
$V^\chi$, which satisfies this constraint, is defined up to the null
state $L_{-1}V_t$
\eqn\RVirnull{L_{-1}V_t=\left(\frac{1}{\hat
k-2}(j^+_{-1}j^-_0+j^-_{-1}j^+_0-2mj^3)+iN\p w+ip\p X\right)V_t}
As a result we are left with one transverse d.o.f.

The $L_1V^\chi=0$ constraint gives
\eqn\RLoVx{a_+(m+j)(m-1-j)+a_-(m-j)(m+1+j)+\frac{iN}{\sqrt{\hk}}
(a_+(m+j)(m-1-j)-a_-(m-j)(m+1+j))=a_X\frac{ip}{2}}
Let us parametrize the solution to this equation by two independent
parameters $a_X,\; a$:
\eqn\RLogensol{a_+=\frac{a_X\frac{ip}{4}+a\left(1-\frac{2m}{\hk}\right)}{(m+j)(m-1-j)}\,,\quad
a_-=\frac{a_X\frac{ip}{4}-a\left(1+\frac{2m}{\hk}\right)}{(m-j)(m+1+j)}}
Therefore the most general massless left-moving state, satisfying
all the Virasoro and $U(1)$ gauge BRST constraints (and defined up
to BRST-exact state $J_{-1}V_t$ and null Virasoro state $L_{-1}V_t$)
is described by holomorphic vertex operator
\eqn\Rmgengf{V^\chi{=}\left(\frac{a_X\frac{ip}{4}{+}a\left
(1{-}\frac{2m}{\hk}\right)}{(m{+}j)(m{-}1{-}j)}j^+_{-1}j^-_0{+}
\frac{a_X\frac{ip}{4}{-}a\left(1{+}\frac{2m}{\hk}\right)}{(m{-}j)
(m{+}1{+}j)}j^-_{-1}j^+_0{+}\frac{4a}{\sqrt{\hk}}\p w{+}a_X\p
X\right)V_t}

Now notice that for
\eqn\Rf{a=ma_1\,,\quad\quad a_X=-i(\hk-2)pa_1}
we obtain that the state \Rmgengf\ is
\eqn\Rfs{V^\chi_0=a_1(-2mJ_{-1}-(\hk-2)L_{-1})V_t}
Such a state is a pure gauge (BRST-exact).

Therefore the most general physical state, which satisfies all the
constraints and which is not a pure gauge, is a state for which
\eqn\Rspsc{\frac{a}{a_X}\neq \frac{im}{(\hk-2)p}}
Any such state is orthogonal to the $V_0^\chi$ state \Rfs, due to
Virasoro and BRST physical state conditions.

The two-point function of the most general physical state \Rmgengf\
is
\eqn\Rgennorm{\langle
V^\chi(p,j,m)V^\chi(-p,j,-m)\rangle=\frac{(\hk^2-\hk
p^2-4m^2)(ma_X+i(\hk-2)pa)^2}{2\hk^2(m^2-j^2)(m^2-(j+1)^2))}\langle
V_t(p,j,m)V_t(-p,j,-m)\rangle}
%
%When \Rspsc\ is satisfied, we are dealing with the norm of the
%physical transverse state.
When \Rspsc\ is not satisfied, we are dealing with the null pure
gauge state, which is a linear combination of timelike and
longitudinal polarizations, that is for such a state
\eqn\Rspsclc{ma_X+i(\hk-2)pa=0}

Finally let us make contact with the result of \ParnachevHH. The two
holomorphic supercurrents of $\NN=2$ supersymmetric $SL(2,R)/U(1)$
gWZW theory are
\eqn\Ntwosusy{G^+=\psi^+ j^-\,,\quad G^-=\psi^-j^+\,.}
Applying the picture-changing operator $G^+_{-1/2}+G^-_{-1/2}$ to
the physical holomorphic vertex operator of \ParnachevHH\ we obtain
the vertex operator of the form
\eqn\pictch{V^\chi=(\p X+a_+j^+_{-1}j^-_0+a_-j^-_{-1}j^+_0+{\rm
fermions})V_t}
Here due to \Rmgengf\ we have
\eqn\apm{a_+=\frac{ip/4}{(m+j)(m-1-j)}\,,\quad
a_-=\frac{ip/4}{(m-j)(m+1+j)}}
Due to \Rgennorm\ we obtain that the two-point function of this
vertex operator has poles at $m=\pm j$. Bellow we discuss these
poles in detail and show that actually only $m=-j$ pole is present,
which after taking into account the mass-shell condition precisely
reproduces the dispersion relation of the gapless low-energy mode,
found in \ParnachevHH.

\subsec{(Anti)holomorphic vertex operators of massless modes in the
$R\times\frac{SL(2,R)\times U(1)}{U(1)}$ coset model}

In this subsection we are going to construct (anti)holomorphic
vertex operators, describing massless (right-)left-moving string
excitations in the $R\times\frac{SL(2,R)\times U(1)}{U(1)}$ model.
The classical geometry of this model is a geometry of the 2d charged
black hole in a direct product with a real line. We choose this real
line as a direction of propagation of all the excitations, and
parametrize it by the coordinate $X$. The momentum of propagation is
$p$.

The vertex operator operators must satisfy BRST and Virasoro
physical state conditions. Recall the null BRST currents \jdef:
\eqn\Rrj{J=\frac{2}{\sqrt{\hk}}j^3\cos\psi+2\sin\psi\p x+2\p w}
\eqn\Rrtj{\tilde J=-\frac{2}{\sqrt{\hk}}\tilde j^3+2\bar\p\tilde w}
Notice that the anti-holomorphic sector is the same as for the model
of the previous subsection: anti-holomorphic (right-moving) sector
of the circle, $U(1)_{\tilde x}$, is disconnected from the rest of
the geometry.

Consider holomorphic sector. Ground state vertex operator is
\eqn\Rrtvo{V_t=e^{ipX}e^{iNw}V_{jm}}
This state must be closed w.r.t. BRST current \Rrj, which imposes
the constraint
\eqn\Rrtachcon{iN=-\frac{2m\cos\psi}{\sqrt{\hk}}}

The most general massless holomorphic vertex operator on the
$R\times\frac{SL(2,R)\times U(1)}{U(1)}$ is given by
\eqn\Rrgfmg{V^\chi=(a_+j^+_{-1}j^-_0+a_-j^-_{-1}j^+_0+a_w\p w+a_X\p
X+b_x\p x +AJ)V_t}
It must be closed w.r.t. BRST current \Rrj, which requires
\eqn\Rrgfbrst{a_w=b_x\sin\psi -\frac{2}{\sqrt{\hk}}\cos\psi
(a_+'-a_-')}
where we have denoted for brevity
\eqn\aprime{a_+'=a_+(m+j)(m-1-j)\,,\quad\quad a_-'=a_-(m-j)(m+1+j)}
The most general massless state, closed w.r.t. \Rrj, is therefore
described by the vertex operator
\eqn\Rrmgbrstcl{V^\chi{=}\left(a_+j^+_{-1}j^-_0{+}a_-j^-_{-1}j^+_0{+}a_X\p
X{+}b_x\p x {+}\left(b_x\sin\psi {-}\frac{2}{\sqrt{\hk}}\cos\psi
(a_+'{-}a_-')\right)\p w{+}AJ\right)V_t}
One d.o.f. in \Rrmgbrstcl\ is unphysical due to the fact that each
state $V^\chi$ is defined up to BRST-exact state $J_{-1}V_t$.
Therefore there remain four d.o.f. of the gauge field $V^\chi$ in
four dimensional target space. Two of them are unphysical, and are
eliminated due to Virasoro constraints, as we show bellow.

Imposing Virasoro constraint $L_1V^\chi=0$, with account to
\Rrtachcon, we obtain condition
\eqn\RrLne{a_+'+a_-'+\frac{2m\cos^2\psi}{\hk}(a_+'-a_-')=a_X\frac{ip}{2}+b_x\frac{m\sin
2\psi}{2\sqrt{\hk}}}
We parametrize the solution to this equation as
\eqn\Rrap{a_+=\frac{a_X\frac{ip}{4}+a\left(1-\frac{2m\cos^2\psi}{\hk}\right)+\frac{b_xm\sin
2\psi}{4\sqrt{\hat k}}}{(m+j)(m-1-j)}}
\eqn\Rram{a_-=\frac{a_X\frac{ip}{4}-a\left(1+\frac{2m\cos^2\psi}{\hk}\right)+\frac{b_xm\sin
2\psi}{4\sqrt{\hat k}}}{(m-j)(m+1+j)}}

Using the mass-shell Virasoro condition (see \Tzvtope)
\eqn\RrmsVc{L_0V^\chi=V^\chi\quad\Rightarrow\quad -\frac{j(j+1)}{\hk
-2}+\frac{p^2-N^2}{4}=0}
we can re-write
\eqn\Rrro{(m+j)(m-1-j)=-\frac{\hk-2}{4}p^2-m\left(1-\frac{2m\cos^2\psi}{\hk}\right)+m^2\sin^2\psi}
\eqn\Rrrt{(m-j)(m+1+j)=-\frac{\hk-2}{4}p^2+m\left(1+\frac{2m\cos^2\psi}{\hk}\right)+m^2\sin^2\psi}
These expressions are useful for computations, described bellow.

To summarize, the most general massless physical state $V^\chi$ on
the $R\times\frac{SL(2,R)\times U(1)}{U(1)}$, satisfying all the
physical constraints, is
\eqn\Rrsgps{V^\chi{=}\left(\frac{a_X\frac{ip}{4}{+}a\left(1{-}\frac{2m\cos^2\psi}{\hk}\right){+}\frac{b_xm\sin
2\psi}{4\sqrt{\hat
k}}}{(m+j)(m-1-j)}j^+_{-1}j^-_0{+}\frac{a_X\frac{ip}{4}{-}a\left(1{+}\frac{2m\cos^2\psi}{\hk}\right){+}\frac{b_xm\sin
2\psi}{4\sqrt{\hat k}}}{(m-j)(m+1+j)}j^-_{-1}j^+_0\right.}
\eqn\Rrsgpst{\left.+a_X\p X+b_x\p x+\left(b_x\sin\psi
-\frac{4a\cos\psi}{\sqrt{\hk}}\right)\p w\right)V_t}
This state is defined up to the null Virasoro state (recall
$n_{L,R}=0$ due to Kaluza-Klein reduction of the $x$-circle)
\eqn\RrVirnull{L_{-1}V'_{jm}=\left(\frac{1}{\hk-2}(j^+_{-1}j^-_0+j^-_{-1}j^+_0-2mj^3)+iN\p
w+ip\p X\right)V_t}
Using \Rrro\ and \Rrrt\ one then demonstrates that for
\eqn\Rrexsol{a_X=-i(\hk-2)pa_1\,,\quad a=ma_1\,,\quad
b_x=-2m\sqrt{\hk}\tan\psi a_1}
the state \Rrsgps\ is non-physical (it is the sum of the BRST-exact
and the null Virasoro states)
\eqn\Rrvxone{V^\chi_0=-a_1\left((\hk-2)L_{-1}+\frac{m\sqrt{\hk}}{\cos\psi}j\right)V_t}

The two-point function of the vertex operator \Rrsgps\ is given by
\eqn\Rrgennorm{\langle V^\chi
V^\chi\rangle{=}\frac{c_1(i(\hk-2)pa{+}ma_X)^2{+}c_2(i(\hk-2)pb_x{-}2m\sqrt{\hk}\tan\psi
a_X)^2{+}c_3(b_x{+}2a\sqrt{\hk}\tan\psi)^2}{(m^2-j^2)(m-(j+1)^2)}\langle
V_tV_t\rangle}
where
\eqn\Rrcone{c_1=\frac{1}{4\hk^2(\hk -2)}(\hk (\hk-2)^2\cos^2\psi
p^2-8(\hk -2)\cos^4\psi m^2-\hk^2(\hk
-2)(p^2-2)+2\hk(\sin^22\psi+2\hk\sin^4\psi)m^2)}
\eqn\Rrctwo{c_2=\frac{\cos^2\psi}{32\hk^2(\hk -2)}(2((\hk -2)^2\cos
2\psi+4-4\hk-\hk^2)m^2+(\hk -2)^2\hk p^2)}
\eqn\Rrthree{c_3=\frac{\cot\psi}{32\hk^2}((8m^2(\hk^2-2m^2)+2(\hk^2-4)m^2p^2-(\hk-2)^2\hk
p^4)\sin 2\psi-m^2(8m^2+(\hk-2)^2p^2)\sin 4\psi )}

When the \Rrexsol\ is satisfied, we are dealing with the null state
$V^\chi_0$ with zero norm.
%When the \Rrexsol\ is not satisfied, we are
%dealing with the physical state $V^X_2$ with positive norm squared.

Like in the previous section, where we derived the vertex operator
\pictch, we now proceed to writing down the vertex operators
$V^x=(\p x+...)V_t$ and $V^X=(\p X+...)V_t$, where dots denote
contribution from $j^\pm _{-1}$ currents:
\eqn\VXc{V^X=\left(\p
X+\frac{ip/4}{(m+j)(m-1-j)}j^+_{-1}j^-_0+\frac{ip/4}{(m-j)(m+1+j)}j^-_{-1}j^+_0\right)V_t}
\eqn\Vxc{V^x=\left(\p
x+\frac{(\sqrt{\hk}/4)\tan\psi}{(m+j)(m-1-j)}j^+_{-1}j^-_0-\frac{(\sqrt{\hk}/4)\tan\psi}{(m-j)(m+1+j)}j^-_{-1}j^+_0\right)V_t}

\newsec{Vertex operators of massless NS-NS states, and correlation functions}

In the previous section we constructed holomorphic and
anti-holomorphic vertex operators, describing respectively
left-moving and right-moving massless excitations of the string on
the $\frac{SL(2,R)\times U(1)_x}{U(1)}$ coset. The state of the
closed string is described by the vertex operator which is a direct
product of holomorphic and anti-holomorphic vertex operators. In
this section we will construct the vertex operators for graviton and
anti-symmetric tensor field, which are massless NS-NS states of
type-II gravity. Kaluza-Klein reduction on $U(1)_x$, applied to
graviton and antisymmetric tensor field vertex operators, gives
vertex operators for gauge fields. We will split the vertex
operators into two decoupled from each other groups, and find
correlation functions for vertex operators within each group.

Denote $M =a,X,x$, and $\mu=a,X$, where $a$ labels non-compactified
directions, transverse to the direction $X$ of propagation of all
the excitations, and $x$ is a coordinate of the compactified circle.
Then, $V^M =j^M V_t$ are holomorphic physical vertex operators and
$\tilde V^M=\tilde j^M V_t$ are anti-holomorphic physical vertex
operators of the massless left-moving and right-moving states.

Here $j^a=\p x^a$ and $\tilde j^a=\bar\p x^a$. Due to \VXc\ and
\Vxc\ the $j^x$ and $j^X$ are elements of two different BRST and
Virasoro cohomology classes, and are defined by
\eqn\jXc{j^X=\p
X+\frac{ip/4}{(m+j)(m-1-j)}j^+_{-1}j^-_0+\frac{ip/4}{(m-j)(m+1+j)}j^-_{-1}j^+_0}
\eqn\jxc{j^x=\p
x+\frac{(\sqrt{\hk}/4)\tan\psi}{(m+j)(m-1-j)}j^+_{-1}j^-_0-\frac{(\sqrt{\hk}/4)\tan\psi}{(m-j)(m+1+j)}j^-_{-1}j^+_0}
Due to anti-holomorphic version of \pictch,
\eqn\tildeJX{\tilde j^X=\bar\p X+\frac{ip/4}{(\bar m+j)(\bar
m-1-j)}\tilde j^+_{-1}\tilde j^-_0+ \frac{ip/4}{(\bar m-j)(\bar
m+1+j)}\tilde j^-_{-1}\tilde j^+_0}
Finally, $\tilde j^x=\bar\p x$.

Notice that due to \Rrgennorm\ the normalized two-point functions
$\langle j^xj^x\rangle _{jm\bar m}/\langle V_tV_t\rangle$, $\langle
j^xj^X\rangle _{jm\bar m}/\langle V_tV_t\rangle$, $\langle
j^Xj^X\rangle _{jm\bar m}/\langle V_tV_t\rangle$ have simple poles
at $m=\pm j$, and due to \Rgennorm\ (the anti-holomorphic version of
it) the two-point function $\langle\tilde j^X\tilde j^X\rangle
_{jm\bar m}/\langle V_tV_t\rangle$ has simple poles at $\bar m=\pm
j$.

The two-point function for ground state of the $SL(2,R)$ model is
given by (see \eg\ \refs{\GiveonPX-\GiveonTQ} for a recent
discussion)
\eqn\Vjmtpf{\langle V_{j,m,\bar m}V_{j,-m,-\bar m}\rangle=\nu
\frac{\Gamma\left(1-\frac{2j+1}{\hat
k-2}\right)\Gamma(-2j-1)\Gamma(1+j+m)\Gamma (1+j-\bar
m)}{\Gamma\left(1+\frac{2j+1}{\hat
k-2}\right)\Gamma(2j+1)\Gamma(-j+m)\Gamma(-\bar m-j)}}
where $\nu$ is some number. Notice that due to factors of
$\Gamma(-j+m)$ and $\Gamma(-\bar m-j)$ in the denominator, the
\Vjmtpf\ has simple zeroth are $j=m$ and $j=-\bar m$. Therefore the
two-point functions $\langle j^xj^x\rangle _{jm\bar m}$, $\langle
j^xj^X\rangle _{jm\bar m}$, $\langle j^Xj^X\rangle _{jm\bar m}$ have
simple pole at $m=-j$, while the simple pole at $m=j$ is canceled,
and the two-point function $\langle\tilde j^X\tilde j^X\rangle
_{jm\bar m}$ has simple pole at $\bar m=j$, while the pole at $\bar
m=-j$ is canceled.

\subsec{Vertex operators and their correlation functions}

Graviton vertex operator is
\eqn\Rrgrvo{G^{MN}=(j^M\tilde j^N +j^N\tilde j^M)V_t\,.}
Antisymmetric tensor field vertex operator is
\eqn\RrBvo{B^{MN}=(j^M\tilde j^N -j^N\tilde j^M)V_t\,.}
Gauge field vertex operators are:
\eqn\GAvo{A^\mu =G^{x\mu}=(j^x\tilde j^\mu+\bar\p xj^\mu)V_t}
\eqn\GAvo{B^\mu =B^{x\mu}=(j^x\tilde j^\mu-\bar\p xj^\mu)V_t}

We have the following groups of vertex operators defined by the spin
w.r.t. to the rotations in the transverse non-compactified space
(for which the coordinates are labeled by small Latin
indices).\foot{See \eg\ \KovtunEV\ for a recent discussion in the
holographic context.} In the sound channel the spin is zero, and one
considers the fields $G^{XX},\;A^X\;B^X$. In the shear channel the
spin is one, and one considers the fields
$G^{Xa},\;B^{Xa},\;A^a,\;B^a$. In the scalar channel the spin is
two, and one considers the fields $G^{ab}$ and $B^{ab}$. Due to the
rotational symmetry in the transverse space, vertex operators from
different groups are decoupled from each other.

{\it Shear channel}

In the shear channel we have vertex operators
\eqn\GXa{G^{Xa}=(j^X\bar\p x^a+\tilde j^X\p x^a)V_t}
\eqn\BXa{B^{Xa}=(j^X\bar\p x^a-\tilde j^X\p x^a)V_t}
\eqn\GAYvo{A^a =G^{xa}=(j^x\bar\p x^a+\bar \p x\p x^a)V_t}
\eqn\GAYvo{B^a =B^{xa}=(j^x\bar \p x^a-\bar \p x\p x^a)V_t}
Notice that all these vertex operators are coupled to each other. We
can consider instead two groups of operators:

the first group is
\eqn\SXa{S^{Xa}=\frac{1}{2}(G^{Xa}+B^{Xa})=j^X\bar\p x^a V_t}
\eqn\RXa{W^{a}=\frac{1}{2}(A^{a}+B^{a})=j^x\bar\p x^a V_t}

and the second group is
\eqn\GAYvo{R^{Xa}=\frac{1}{2}(G^{Xa}-B^{Xa})=\tilde j^X\p x^aV_t}
\eqn\GAYvo{U^{a}=\frac{1}{2}(A^{a}-B^{a})=\bar \p x\p x^a V_t}

We call the operators from the first group S-system and the
operators from the second group R-system. The S-system is decoupled
from the R-system. For the vertex operators of the S-system the
two-point functions are
\eqn\SaSacor{\langle S^{Xa}S^{Xb}\rangle
=-\frac{1}{2}\delta^{ab}\langle j^Xj^X\rangle _{jm\bar m}}
\eqn\WaWacor{\langle W^{a}W^{b}\rangle
=-\frac{1}{2}\delta^{ab}\langle j^xj^x\rangle _{jm\bar m}}
\eqn\SaWacor{\langle S^{Xa}W^{b}\rangle
=-\frac{1}{2}\delta^{ab}\langle j^xj^X\rangle _{jm\bar m}}
These correlation functions have a simple pole at $j=-m$.

For the vertex operators of the R-system the two-point functions are
\eqn\RaRacor{\langle R^{Xa}R^{Xb}\rangle
=-\frac{1}{2}\delta^{ab}\langle\tilde j^X\tilde j^X\rangle _{jm\bar
m}}
\eqn\UaUacor{\langle U^{a}U^{b}\rangle =\frac{1}{4}\delta^{ab}}
\eqn\RaUacor{\langle R^{Xa}U^{b}\rangle =0}
These correlation functions have a simple pole at $j=\bar m$.

Due to holographic correspondence we obtain correlation functions of
the shear components of the stress-energy tensor of the dual field
theory:\foot{One also may be interested in computing correlation
functions of the operator, dual to $B_{MN}$-field. See
\refs{\DasEI,\FerraraBP}, where the primary operator in $\NN=4$ SYM,
holographically dual to the $B$-field in $AdS_5\times S^5$, was
found.}
\eqn\Txacore{\langle G^{Xa}G^{Xb}\rangle=\langle
T^{Xa}T^{Xb}\rangle=-\frac{1}{2}\delta^{ab}\left(\langle
j^Xj^X\rangle _{jm\bar m}+\langle\tilde j^X\tilde j^X\rangle
_{jm\bar m}\right)}
The correlation functions for the transverse components of the
charge current are
\eqn\jXjX{\langle J^aJ^b\rangle=\langle A^aA^b\rangle
=-\frac{1}{2}\delta^{ab}\left(\langle j^xj^x\rangle_{jm\bar m}
-\frac{1}{2}\right)}
Finally,
\eqn\jXTXa{\langle J^aT^{Xb}\rangle =\langle
A^aG^{Xb}\rangle=-\frac{1}{2}\delta^{ab}\langle j^xj^X\rangle
_{jm\bar m}}
We conclude that in the shear/transverse diffusion channel we have
modes with the dispersion relations $m=-j$ and $\bar m=j$.

{\it Sound channel}

In the sound channel we have vertex operators
\eqn\GXX{G^{XX}=j^X\tilde j^XV_t}
\eqn\AX{A^{X}=G^{xX}=(j^x\tilde j^X+\bar \p x j^X)V_t}
\eqn\BX{B^{X}=B^{xX}=(j^x\tilde j^X-\bar \p x j^X)V_t}
Notice that $A^X$ and $B^X$ are coupled. Consider instead decoupled
gauge fields vertex operators
\eqn\WX{W^{X}=\frac{1}{2}(A^X+B^X)=j^x\tilde j^XV_t}
\eqn\UX{U^{X}=\frac{1}{2}(A^X-B^X)=\bar \p x j^XV_t}

Correlation functions are
\eqn\GXXcor{\langle G^{XX}G^{XX}\rangle =\langle j^X j^X\rangle
_{jm\bar m}\langle\tilde j^X\tilde j^X\rangle _{jm\bar m}}
\eqn\GXWXcor{\langle G^{XX}W^{X}\rangle =\langle j^x j^X\rangle
_{jm\bar m}\langle\tilde j^X\tilde j^X\rangle _{jm\bar m}}
\eqn\GXUXcor{\langle G^{XX}U^{X}\rangle =0}
\eqn\WXcor{\langle W^{X}W^{X}\rangle =\langle j^x j^x\rangle
_{jm\bar m}\langle \tilde j^X \tilde j^X\rangle _{jm\bar m} }
\eqn\UXcor{\langle U^{X}U^{X}\rangle =-\frac{1}{2}\langle j^X
j^X\rangle _{jm\bar m}}
The correlation functions \GXXcor, \GXWXcor\ and \WXcor\ have simple
poles at $j=-m$ and $j=\bar m$ and the correlation function \UXcor\
has simple pole at $j=-m$.

Due to holographic correspondence we obtain correlation functions of
the longitudinal component of the stress-energy tensor of the dual
field theory:
\eqn\TXXbcore{\langle T^{XX}T^{XX}\rangle =\langle
G^{XX}G^{XX}\rangle =\langle j^X j^X\rangle _{jm\bar m}\langle\tilde
j^X\tilde j^X\rangle _{jm\bar m}}
The correlation function of the longitudinal component of the charge
current is
\eqn\JXJX{\langle J^XJ^X\rangle=\langle A^XA^X\rangle =\langle
j^xj^x\rangle_{jm\bar m}\langle\tilde j^X\tilde j^X\rangle_{jm\bar
m} -\frac{1}{2}\langle j^Xj^X\rangle_{jm\bar m}}
Finally,
\eqn\JXTXa{\langle J^XT^{XX}\rangle =\langle
A^XG^{XX}\rangle=\langle j^xj^X\rangle _{jm\bar m}\langle\tilde
j^X\tilde j^X\rangle _{jm\bar m}}
Therefore in the sound channel we have modes with the dispersion
relations $m=-j$ and $\bar m=j$.

{\it Scalar channel}

In the scalar channel one considers $G^{ab}$ and $B^{ab}$,
correlation functions for which do not have poles at $j=-m$ and
$j=\bar m$.

\subsec{Low-energy modes}

In the previous subsection we concluded that there are modes with
the dispersion relations $m=-j$ and $\bar m=j$ in the shear and
sound channels of the quantum field theory holographically dual to
the charged black brane. Now we are going to show, considering small
$\omega$ and $q$, that these modes are actually gapless modes.

The frequency is determined by the asymptotic behavior of the
tachyon vertex operator $V_t\sim e^{i\omega t}$ (see \GiveonGE), and
is given by (for $\psi\neq \pi/2$, that is in non-extremal case)
\eqn\omegd{\omega=\frac{(1-\tan^2(\psi/2))(m-\bar m)}{\sqrt{\hat
k}}}
Due to the gauge physical state condition \pscg\ we have $\bar
m=-m\cos\psi$, and therefore (also after Wick rotation $m\rightarrow
im$)
\eqn\ommm{\omega=\frac{2im\cos\psi}{\sqrt{\hat k}}}
Due to the mass-shell condition
\eqn\RmsVc{-\frac{j(j+1)}{\hk}+\frac{p^2-N^2}{4}=0}
where  $N\sim m\sim \omega$, for $\omega\sim p^2$ and $p\ll 1$ we
obtain
\eqn\jinfp{j=\frac{\hat k}{4}p^2}

Therefore the S-system possesses the low-energy excitation mode with
the dispersion relation
\eqn\jeqm{\omega=-i\frac{\sqrt{\hat k}}{2} \cos\psi\, p^2}
while for the R-system we obtain the mode with the dispersion
relation
\eqn\jeqmc{\omega=-i\frac{\sqrt{\hat k}}{2} \, p^2}

In the extremal case $\psi=\pi/2$ the two-point functions in the
S-system, due to \jeqm, behave as $\langle SS\rangle\sim 1/\omega$,
indicating local criticality, while the dispersion relation \jeqmc\
of the R-system stays unaffected.

At zero density $q=0$ we have $\psi =0$, and the mode \jeqm\
coincides with the mode \jeqmc. In this case $U(1)_x$ completely
decouples from $SL(2,R)/U(1)$, and we recover the results of
\ParnachevHH\ for the model on the $SL(2,R)/U(1)$. Due to \betab\ we
obtain
\eqn\visczd{\omega=-i\frac{1}{4\pi T}p^2}
Comparing it with the shear mode dispersion relation at zero density
\eqn\hydr{\omega =-i\frac{\eta}{sT}p^2}
we recover the result of \ParnachevHH\
\eqn\etasqzero{\frac{\eta}{s}=\frac{1}{4\pi}\,.}
%

%Finally, in the extremal case, $\psi =\pi/2$, the background metric
%is \GiveonZZ
%%
%\eqn\bmextrc{ds^2=\frac{\hk}{4} (d\chi^2-\sinh^2\chi  dt^2)}
%%
%This is metric of $AdS_2$ with radius $R_2=\sqrt{\hk}/2$. Due to
%physical state condition \pscg\ we have $\bar m=0$. Therefore $\bar
%m=j$ pole gives rise to $p=0$ pole. Due to \GiveonGE\ the
%near-boundary behavior of $V_t$ is given by $V_t\simeq e^{imt}$. The
%frequency is therefore (also after Wick rotation $m\rightarrow im$)
%given by $\omega =im/R_2=2im/\sqrt{\hk}$. Therefore the $m=-j$
%dispersion relation can be rewritten as
%%
%\eqn\extrpole{\omega =-i\frac{\sqrt{\hk}}{2}p^2\,,}
%%
%which coincides with the dispersion relation \jeqmc.

\subsec{``$2k_F$" singularity}

Expression \Vjmtpf\ for the groundstate two-point function contains
the factor of $\Gamma \left(1-\frac{2j+1}{\hat k-2}\right)$. Due to
the physical state mass-shell condition \RmsVc\ we obtain
\eqn\jexprN{j=-\frac{1}{2}+\frac{1}{2}\sqrt{1+(\hat k-2)(p^2-N^2)}}
At zero frequency $N=0$. Therefore equation $2j+1=\hat k-2$, which
defines singularity of $\Gamma \left(1-\frac{2j+1}{\hat
k-2}\right)$, has a zero frequency and finite momentum solution. The
value of the momentum is given by
\eqn\pstar{p_\star^2=\frac{1}{\ell_s^2}\left(\hat k-2-\frac{1}{\hat
k-2}\right)}
Note that $p_*$ is independent of the chemical potential
$\mu=\tan\frac{\psi}{2}$. The singular behavior of $\langle
V_{jm\bar m}V_{jm\bar m}\rangle$ at $\omega =0$ and $p=p_*$ was
compared by Polchinski and Silverstein \PolchinskiNH\ with ``$2k_F$"
singularities in current correlation functions of condensed matter
systems (see \eg\ \GiulianiV).

\newsec{Type-II gravity approximation}

In this section we will compute the two-point functions for
graviton, antisymmetric tensor field and gauge fields in the
background of the 2d charged black hole (in a direct product with a
flat space). Our purpose is to verify the dispersion relations
\jeqm-\jeqmc.

One-loop beta-functions for the NS-NS fields of type-II gravity are
given by (see \eg\ \refs{\PolchinskiRQ,\PolchinskiRR})
\eqn\gmneq{\beta^G_{MN}=R_{MN}+2\nabla_M\p_N\Phi -\frac{1}{4}H_M
^{\;\;LS}H_{NLS}\,,}
\eqn\dileq{\beta^\Phi=c+\frac{1}{16\pi ^2}\left(4
(\p\Phi)^2-4\nabla^2\Phi -R+\frac{1}{12}H^2\right)\,,}
\eqn\Amueq{\beta^B_{MN}=\nabla_L
H^L_{\;\;MN}-2(\p_L\Phi)H^L_{\;\;MN}}
Corresponding equations of motion are $\beta^{G,\Phi,B}=0$.

Here the field strength of antisymmetric tensor $B_{MN}$ is given by
\eqn\htensor{H_{MNL}=\p_M B_{NL}+\p_N B_{LM}+\p_L B_{MN}}
The beta-functions \gmneq-\Amueq\ are invariant w.r.t. the gauge
symmetry
\eqn\Bmngauge{\delta B_{MN}=\p_M \Lambda _N -\p_N \Lambda_M\,.}

Requirement of the worldsheet conformal invariance gives the
equations of motion $\beta^{G,B,\Phi}=0$. These equations have a
black brane solution, which is a direct product of two-dimensional
charged black hole (CBH) and flat space, $CBH\times R^{d-1}$:
\eqn\solgr{g_{MN}={\rm diag}\{-f(r),1/f(r),1,...,1\}\,,\quad
f(r)=1-2Me^{-Qr}+q^2e^{-2Qr}}
\eqn\solgrt{\Phi=\Phi_0-\frac{Qr}{2}\,,\quad\quad
F_{tr}=F(r)=Qqe^{-Qr}\,.}
where\foot{We thank A. Giveon for pointing out to us the role of
this equation in the 2d charged black hole solution of type-II
superstring theory.}
\eqn\AtKK{g_{tx}=B_{xt}=-B_{tx}=A_t}
The string theory solution, described in the previous section,
implies $Q=2/\sqrt{\hk}$.

Consider fluctuations $h_{MN}$, $b_{MN}$ and $\varphi$ around this
solution. Use the diffeomorphism invariance to fix $h_{M r}=0$. Use
gauge invariance \Bmngauge\ to fix $b_{M r}=0$. Among $d+1$
space-time coordinates, denoted by capital Latin indices, we have
$t,r$ coordinates of the charged black hole and $d-1$ flat
coordinates. Let us consider $CBH\times R^3$. Choose $X$ to be the
$R^3$ direction of propagation of excitations (with momentum $p$)
and choose $Y$ to be the $R^3$ direction, transverse to propagation
of excitations. Finally $x$ is the direction of $R^3$ which we are
going to Kaluza-Klein reduce. Fluctuations depend on $t,r,X$. The
dependence on $t$ and $X$ in momentum representation boils down to
the factor $e^{-i\omega t+ipX}$.

Perform Kaluza-Klein reduction of the $x$ coordinate. Small Greek
indices are used for non-reduced coordinates, $M=\mu, x$. It is
convenient, as we did in the worldsheet consideration, to consider
fluctuations of the fields
\eqn\Smn{S_{MN}=\frac{1}{2}(h_{MN}+b_{MN})}
\eqn\Rmn{R_{MN}=\frac{1}{2}(h_{MN}-b_{MN})}
The fields \Smn\ belong to the S-system and the fields \Rmn\ belong
to the R-system, we are using the same terminology as in the
previous section.

Let us consider shear fluctuations in the reduced space $CBH\times
R^2$: $R_{tY}$, $R_{XY}$, $S_{tY}$, $S_{XY}$ and transverse
components of gauge fields $w_Y$ and $u_Y$ (see bellow). The ansatz
for graviton and two-form field in the non-reduced space $CBH\times
R^3$ in terms of the fields on the reduced space $CBH\times R^2$ is
\eqn\GMNAnz{G=\left({{A_t^2-f\atop{0\atop
0}}\atop{R_{tY}+S_{tY}+A_t(u_Y+w_Y)\atop A_t}}\;{{0\atop{1/f\atop
0}}\atop{0\atop 0}}\;{{0\atop{0\atop 1}}\atop{R_{XY}+S_{XY}\atop
0}}\;{{R_{tY}+S_{tY}+A_t(u_Y+w_Y)\atop {0\atop
R_{XY}+S_{XY}}}\atop{1\atop u_Y+w_Y}}\;{{A_t\atop {0\atop
0}}\atop{u_Y+w_Y\atop 1}}\right)}
\eqn\BMNAnz{B=\left({{0\atop{0\atop
0}}\atop{-(S_{tY}-R_{tY})-A_t(w_Y-u_Y)\atop A_t}}\;{{0\atop{0\atop
0}}\atop{0\atop 0}}\;{{0\atop{0\atop 0}}\atop{-(S_{XY}-R_{XY})\atop
0}}\;{{S_{tY}-R_{tY}+A_t(w_Y-u_Y)\atop {0\atop
S_{XY}-R_{XY}}}\atop{0\atop w_Y-u_Y}}\;{{-A_t\atop {0\atop
0}}\atop{-(w_Y-u_Y)\atop 0}}\right)}

Before proceeding, rescale
\eqn\Qres{\r =rQ\,,\quad\w=\omega/Q\,,\quad \pp=p/Q\,,}
which eliminates $Q$ dependence from the equations of motion.

Due to string theory result we know that the R-system fields $R$ and
$u$ are decoupled from the S-system fields $S$ and $w$. We will find
out that this decoupling is true in gravity computations as well. To
find equations of motion for the fields $R_{\mu\nu}$, $S_{\mu\nu}$,
$w_\mu$ and $u_\mu$ we compute the beta functions $\beta
^R_{MN}=\beta ^G_{MN}-\beta ^B_{MN}$ and $\beta ^S_{MN}=\beta
^G_{MN}+\beta ^B_{MN}$ for $(MN)=(tY),\; (rY),\;(XY),\;(xY)$.

Consider first equations of motion in the R-system.

$\beta ^R_{tY}$:
\eqn\betaRtY{pf (pR_{tY}+\omega R_{XY})+f^2(2\Phi'
R_{tY}'-R_{tY}'')-}
\eqn\betaRtYt{-A_t(f^2u_Y''+f(f'-2f\Phi')u_Y'+(\omega^2-p^2f+(f^2/A_t)(A_t''-2A_t'\Phi')u_Y)=0}

$\beta ^R_{rY}$:
\eqn\betaRrY{\omega R_{tY}'+pfR_{XY}'=0}

$\beta^R_{XY}$:
\eqn\betaRXY{\omega (pR_{tY}+\omega
R_{XY})+f(fR_{XY}''+(f'-2f\Phi')R_{XY}')=0}

$\beta^R_{xY}$:
\eqn\betaRxY{f^2u_Y''+f(f'-2f\Phi')u_Y'+(\omega^2-p^2f)u_Y=0}

Notice that for the CBH background $\Phi'=-1/2$ and $A_t''=-A_t'$.
Therefore one sees that $u_Y$ contribution to $R_{tY}$ equation
\betaRtY\ vanishes due to $u_Y$ equation \betaRxY. Therefore we see
that $R_{\mu Y}$ and $u_Y$ fluctuations decouple, as expected from
the string theory computations \RaUacor.

Introduce diff-invariant quantity
\eqn\Zdef{Z=pR_{tY}+\omega R_{XY}}
Solving following from this definition equation
\eqn\Zprdef{Z'=pR_{tY}'+\omega R_{XY}'}
together with $R_{rY}$ equation \betaRrY\ one obtains
\eqn\RtYXYpr{R_{tY}'=-\frac{pfZ'}{\omega^2-p^2f}\,,\quad\quad
R_{XY}'=\frac{\omega Z'}{\omega^2-p^2f}}
Plugging expressions \RtYXYpr\ into $R_{tY}$ equation \betaRtY\ one
obtains equation (the same equation is obtained if one plugs
\RtYXYpr\ into $R_{XY}$ equation \betaRXY)
\eqn\Zeq{Z''+\left(\frac{\omega^2
f'}{f(\omega^2-p^2f)}-2\Phi'\right)Z'+\frac{\omega^2-p^2f}{f^2}Z=0}
Together with decoupled from it transverse gauge field equation
\betaRxY\ for $u_Y$ these are fluctuation equations for shear
components of R-system.

Consider now $S$ and $w$ fluctuation equations of the S-system.
$\beta^S_{tY}$:
\eqn\betaStY{-\omega^2 A_t w_Y+f(p(p S_{tY}+\omega
S_{XY})+A_t(w_Y(p^2-2A_t^{\prime 2})-2A_t'S_{tY}'-f'w_Y'))-}
\eqn\betasTyt{-f^2(-2\Phi '(S_{tY}'+A_tw_Y')+2A_t'(w_Y'-\Phi'
w_Y)+A_t''w_Y+S_{tY}''+A_tw_Y'')=0}

$\beta^S_{rY}$:
\eqn\betaSrY{2\omega A_t' w_Y+\omega S_{tY}'+pf S_{XY}'=0}

$\beta^S_{XY}$:
\eqn\betaSXY{\omega (pS_{tY}+\omega S_{XY})+f(S_{XY}'(f'-2\Phi
'f)+fS_{XY}'')=0}

$\beta^S_{xY}$:
\eqn\betaSxY{(\omega^2-f(p^2-2A_t^{\prime
2}))w_Y+f(2A_t'S_{tY}'+(f'-2\Phi 'f)w_Y'+fw_Y'')=0}

Introduce diff-invariant quantity
\eqn\Vdef{V=pS_{tY}+\omega S_{XY}}
Then solving equation
\eqn\Vprime{V'=pS_{tY}'+\omega S_{XY}'}
together with $\beta^S_{rY}$ equation \betaSrY\ we obtain
\eqn\SXYtypr{S_{XY}'=\frac{\omega
(2pA_t'w_Y+V')}{\omega^2-p^2f}\,,\quad S_{tY}'=-\frac{2\omega
^2A_t'w_Y+pfV'}{\omega^2-p^2f}}
Plugging into $\beta^S_{tY}$ equation \betaStY\ the expressions
\SXYtypr\ together with $w_Y''$, expressed from $w_Y$ equation
\betaSxY, we arrive at (the same result is obtained by plugging
\SXYtypr\ into $\beta^S_{XY}$ equation \betaSXY) \foot{Also take
into account $\Phi'=-1/2$ and $A_t''=-A_t'$.}
\eqn\Vfleq{V''+\left(\frac{\omega^2
f'}{f(\omega^2-p^2f)}-2\Phi'\right)V'+\frac{\omega^2-p^2f}{f^2}V+
\frac{2pA_t'}{f}\left(fw_Y'+\frac{\omega^2f'}{\omega^2-p^2f}w_Y\right)=0}
Finally, using $S_{tY}'$, expressed in \SXYtypr, in $w_Y$ equation
\betaSxY\ we obtain
\eqn\wYfleq{w_Y''+\frac{f'-2f\Phi'}{f}w_Y'+\left(\frac{\omega^2-p^2f}{f^2}+\frac{2A_t^{\prime
2}}{f}\left(1-\frac{2\omega^2}{\omega^2-p^2f}\right)\right)w_Y-\frac{2pA_t'}{\omega^2-p^2f}V'=0}
We see that in the S-system tensor field shear components are
coupled to gauge field transverse component, which agrees with
string computation \SaWacor.

Let us look for poles of the correlation functions in R-system and
in S-system. Notice that R-system is just S-system at vanishing
background flux $A_t'=0$; compare equation \Vfleq\ with equation
\Zeq\ and equation \wYfleq\ with equation \betaRxY\ to see that.
Therefore it is sufficient to study S-system.

Introduce new radial coordinate $u=e^\r$. Then inner and outer
horizons are located at
\eqn\upm{u_\pm =M\pm\sqrt{M^2-q^2}}
The equations of motion \Vfleq\ and \wYfleq\ become (also take into
account $q=\sqrt{u_+u_-}$)
\eqn\aYu{\frac{d^2w_Y}{du^2}+\left(\frac{1}{u-u_-}+\frac{1}{u-u_+}\right)
\frac{dw_Y}{du}+\frac{1}{(u-u_-)(u-u_+)}\left(\frac{\w^2u^2-\pp^2(u-u_-)(u-u_+)}{(u-u_-)(u-u_+)}+\right.}
\eqn\aYu{\left.+\frac{2u_+u_-}{u^2}-\frac{2u_+u_-\w^2}{\w^2u^2-\pp^2(u-u_-)(u-u_+)}\right)w_Y+
\frac{2\pp\sqrt{u_+u_-}}{\w^2
u^2-\pp^2(u-u_-)(u-u_+)}\frac{d\V}{du}=0}
\eqn\Zu{\frac{d^2\V}{du^2}+\frac{1}{u}\left(2+\frac{\w^2 u^2}{\w^2
u^2-\pp^2(u-u_-)(u-u_+)}\left(\frac{u_+}{u-u_+}+\frac{u_-}{u-u_-}\right)\right)
\frac{d\V}{du}-\frac{2\pp\sqrt{u_+u_-}}{u^2}\frac{dw_Y}{du}+}
\eqn\Zu{+\frac{\w^2u^2{-}\pp^2(u-u_-)(u-u_+)}{(u-u_-)^2(u-u_+)^2}\V{-}\frac{2\pp\sqrt{u_+u_-}}{u}
\frac{\w^2}{\w^2u^2{-}\pp^2(u-u_-)(u-u_+)}\left(\frac{u_+}{u-u_+}{+}
\frac{u_-}{u-u_-}\right)w_Y{=}0}

In the near horizon limit $v=u-u_+\ll 1$ equations \aYu\ and \Zu\
give rise to
\eqn\qYunh{\frac{d^2w_Y}{dv^2}+\frac{1}{v}\frac{dw_Y}{dv}+\frac{\w^2u_+^2}
{(u_+-u_-)^2v^2}w_Y=0}
\eqn\Zunh{\frac{d^2\V}{dv^2}+\frac{1}{v}\frac{d\V}{dv}+\frac{\w^2u_+^2}
{(u_+-u_-)^2v^2}\V=0}
The incoming-wave solutions are
\eqn\aYuincw{w_Y(u)=C_1(u-u_+)^\frac{-i\w u_+}{u_+-u_-}\,,\quad
\V(u)=C_2(u-u_+)^\frac{-i\w u_+}{u_+-u_-}}

In the asymptotic region $u\gg 1$ equations \aYu\ and \Zu\ give rise
to
\eqn\aYnb{\frac{d^2w_Y}{du^2}+\frac{2}{u}\frac{dw_Y}{du}+\frac{\w^2-\pp^2}{u^2}w_Y=0}
\eqn\ZYnb{\frac{d^2\V}{du^2}+\frac{2}{u}\frac{d\V}{du}+\frac{\w^2-\pp^2}{u^2}\V=0}
with the solution
\eqn\aYnbsol{w_Y=\AA _w
u^{\frac{1}{2}(-1+\sqrt{1+4(\pp^2-\w^2)})}+\BB_w
u^{\frac{1}{2}(-1-\sqrt{1+4(\pp^2-\w^2)})}}
\eqn\ZYnbsol{\V=\AA_V
u^{\frac{1}{2}(-1+\sqrt{1+4(\pp^2-\w^2)})}+\BB_V
u^{\frac{1}{2}(-1-\sqrt{1+4(\pp^2-\w^2)})}}

We solve numerically the equations \aYu\ \Zu\ with boundary
conditions \aYuincw\ and find two linearly-independent solutions
$(w_Y^{(1)},\V^{(1)})$ and $(w_Y^{(2)},\V^{(2)})$ (for two
independent choices of $C_{1,2}$). The correlation matrix is given
by \KaminskiDH\ $\GG\simeq \BB \AA^{-1}$, where the matrices of
leading and subleading coefficients are determined by \aYnbsol\ and
\ZYnbsol:
\eqn\AABB{\AA=\left({\AA_V^{(1)}\atop
\AA_w^{(1)}}\;{\AA_V^{(2)}\atop \AA_w^{(2)}}\right)\,,\quad
\BB=\left({\BB_V^{(1)}\atop \BB_w^{(1)}}\;{\BB_V^{(2)}\atop
\BB_w^{(2)}}\right)}
Zeroes of the determinant of the matrix of the leading behavior
coefficients \foot{See \eg\ \GoykhmanVY, where computation of
correlation matrix in the different system of two coupled
differential equations is explained in detail.}
\eqn\zerodet{\AA _V^{(1)}\AA _w^{(2)}-\AA _V^{(2)}\AA _w^{(1)}}
define the dispersion relation of low-energy mode, which is given by
\eqn\zdet{\w=-i\pp^2\cos\psi}
Due to \Qres\ and $Q=2/\sqrt{\hk}$ the dispersion relation \zdet\
coincides with the dispersion relation \jeqm, obtained for the
S-system by the worldsheet computation.

From the S-system result \zdet\ we conclude that in the R-system
$\langle ZZ\rangle $ correlation function has pole at
\eqn\zdetzch{\w=-i\pp^2\,,}
which coincides with the pole \jeqmc, obtained by the worldsheet
computation. Notice that as in \ParnachevHH\ the supergravity result
does not receive stringy corrections.

\newsec{Discussion}

In this paper we have used the holographically dual string theory to
study quantum field theory at finite temperature and chemical
potential. The string theory was defined by the gWZW model on the
$\frac{SL(2,R)\times U(1)_x}{U(1)}\times R^{d-1}$, with the $U(1)$
gauged asymmetrically \GiveonGE. Using BRST quantization on the
coset and the covariant quantization of the string, we have
constructed vertex operators, representing massless NS-NS states of
the string. The gauge fields vertex operators were obtained by the
Kaluza-Klein reduction of the graviton and the two-form field vertex
operators on the $U(1)_x$.

We have found that these vertex operators split into two decoupled
systems. This implies that the boundary low energy theory splits
into two decoupled models, as far as the two-point functions are
concerned. At low energies the Green's functions of stress energy
tensor and global $U(1)$ current exhibit two gapless poles.
Corresponding  dispersion relations are \jeqm\  and \jeqmc\  in the
shear and sound channels. The dispersion relation \jeqmc\ does not
depend on the charge to mass ratio of the charged black hole
background. When the charge density is zero, the dispersion relation
\jeqm\ coincides with the dispersion relation \jeqmc. We have
verified these results by computations in type-II supergravity; the
supergravity results exactly coincide with superstring results. We
speculate that the system is described at low energies by a
decoupled sum of two non-interacting fluids. It would be interesting
to make this picture more precise.

The current and stress-energy tensor two-point correlation
functions, which we have computed, also possess finite-momentum
zero-frequency singularity. As in \PolchinskiNH\ it originates from
the two-point function of the vertex operator of the ground state of
the WZW model on $SL(2,R)$. This ``$2k_F$" singularity is a purely
stringy effect \PolchinskiNH, absent in the supergravity
approximation: from \pstar\ it follows, that the momentum $p_*$,
measured in units of inverse curvature radius, scales as
$(Rp_*)^2\simeq \hat k\ell_s^2p_*^2\sim \hat k^2$ when $\hat k$ is
large. Therefore in supergravtiy approximation $p_*$ is
parametrically large.

We have also studied the shear channel in heterotic gravity (see
Appendix B), and found one low-energy mode. Matching its dispersion
relation to the one obtained from the thermodynamics of the 2d
charged black hole, we have derived $\eta/s=1/(4\pi)$ for any charge
to mass ratio. It would be interesting to obtain this result from
heterotic string theory as well. However naive construction of the
heterotic string theory, based on the $\frac{SL(2,R)\times
U(1)_x}{U(1)}$ coset model (where $U(1)_x$ is holomorphic, that is a
part of internal space from purely bosonic left-moving sector of
heterotic string theory), appears to contain $U(1)$ chiral anomaly.
Indeed, naively, to construct heterotic string, based on the coset
model used in this paper, one takes the gWZW action \gWZWkk\ and
adds to it the Dirac term $S_f\simeq \int d^2z\,{\rm
Tr}\,\tilde\Psi(\p +A)\tilde\Psi$, where anti-holomorphic
(right-moving) fermions $\tilde\Psi\in sl(2,R)\ominus u(1)$ are
superpartners of the anti-holomorphic bosonic currents on
$SL(2,R)/U(1)$, and $A$ is the $U(1)$ gauge field. Due to such
chiral interaction, on the quantum level the anomaly appears, and
the theory becomes inconsistent.

This issue was actually resolved in a different heterotic coset
stringy realization of the 2d charged black hole \JohnsonJW. As it
was observed there, the chiral anomaly due to fermions should be
compensated by the classical anomaly of gWZW action for bosons
\WittenMM. In fact, bosonization of the fermions results in the
chiral anomaly due to fermions appearing on the classical level,
just as in the anomalous gWZW action \JohnsonKV. Therefore
separately the bosonic and fermionic parts of the action are not
invariant under $U(1)$ gauge transformation, while their sum is
invariant. It is not clear however how these ideas can be directly
applied to the model, based on the bosonic action \gWZWkk, which was
constructed \GiveonGE\ to be anomaly-free on its own.
\bigskip

{\bf Acknowledgements:} We thank J. de Boer, B.~Galilo, A.~Giveon,
M.~Kulaxizi, D.~Kutasov, S.~Sidorov, E.~Verlinde and J.~Zaanen for
discussions. We are grateful to the \'Ecole Normale Sup\'erieure,
where part of this work was completed, for hospitality. This work
was supported in part by a VIDI innovative research grant from NWO.

\appendix{A}{Conventions and review of the gWZW model on the
$\frac{SL(2,R)\times U(1)}{U(1)}$}

%\noindent

\subsec{Conventions}

Put the string length equal to one, $\alpha
'\equiv\frac{\ell_s^2}{2}=\frac{1}{2}$. The contribution to the
worldsheet stress-energy tensor, coming from the coordinates $X^\mu
(z,\bar z)$ of the flat subspace of the target space-time, is given
by
\eqn\Tflat{T_{flat}(z)=-\p X^\mu (z)\p X_\mu (z)}
and similarly for the anti-holomorphic part $\tilde T(\bar z)$. The
Polyakov action is
\eqn\flact{S_P=\frac{1}{2\pi}\int d^2z\,\p X\bar\p X\,.}
%
%The ground state vertex operator for the mode with momentum $p$ is
%%
%\eqn\Vflat{V_{flat}=e^{ip\cdot X(w,\bar w)}}
%%
The two-point function is
\eqn\Xfcor{\langle X^\mu (z,\bar z)X^\nu (w,\bar w)\rangle
=-\frac{1}{2}\eta^{\mu\nu}\left(\log (z-w)+\log(\bar z-\bar
w)\right)}
%
%we obtain
%%
%\eqn\TVflOPE{T_{flat}(z) V_{flat}(w,\bar
%w)=\frac{1}{4}\frac{p^2}{(z-w)^2}V_{flat}(w,\bar w)+\cdots\,,}
%%
%and therefore scaling dimension of the operator $V_{flat}$ is equal
%to
%%
%\eqn\FlXscd{\Delta_{flat}=\frac{k^2}{4}\,.}
%%

The Kac-Moody
%symmetry
%transformations in WZW model are given by
%%
%\eqn\gtransf{g(z,\bar z)\rightarrow g'(z,\bar z)=\Omega (z)g(z,\bar
%z)\tilde\Omega ^{-1}(\bar z)\,.}
%%
%Corresponding
holomorphic (left-moving) and anti-holomorphic (right-moving)
currents of the WZW model at level $\hat k$ are given by
\eqn\curdef{j(z)=j_At^A=-\frac{\hat k}{2}\p gg^{-1}\,,\quad\quad
\tilde j(\bar z)=\tilde j_At^A=\frac{\hat k}{2}g^{-1}\bar\p g\,.}
Here hermitean generators of a gauge algebra are
\eqn\hgs{t^A=j_0^A\,,\quad\quad [t^A,t^B]=if^{ABC}t^C\,.}
For $SL(2,R)$, which is the group we are interested in, the
following expressions in terms of Pauli matrices take place:
\eqn\sltrg{j^A_0=\frac{1}{2}\sigma ^A\,,\quad\quad
f^{ABC}=\epsilon^{ABC}\,,}
and indices are raised and lowered with the help of $\eta_{AB}={\rm
diag}\{1,1,-1\}$. In Euclidean realization of $SL(2,R)$ we put
$\eta^{AB}=\delta^{AB}$.

The holomorphic currents of Kac-Moody algebra satisfy the following
OPE
\eqn\jjope{j^A(z)j^B(w)=\frac{\frac{ \hat
k}{2}\eta^{AB}}{(z-w)^2}+\frac{if^{ABC}}{z-w}j_C(w)\,,}
and similarly for the anti-holomorphic currents.

%Define the index in adjoint representation
%%
%\eqn\cvff{c_V\eta^{AB}=f^{ACD}f^{B}_{\;\;CD}\,.}
%%
%Then
The holomorphic component of the stress-energy is given by the
Sugawara expression
\eqn\TjjS{T(z)=\frac{1}{\kappa}\eta_{AB}j^A(z)j^B(z)\,,}
similar expression is true for the anti-holomorphic component. Here
\eqn\kapkcv{\kappa=\hat k+c_V\,.}
For $SU(2)$ (and for Euclidean $SL(2,R)$) the index of the adjoint
representation is $c_V=2$ and for $SL(2,R)$ it is $c_V=-2$.

The groundstate representation space of the $SL(2,R)$ currents is
formed by the primary fields $V_j(x,\bar x;w,\bar w)$, characterized
by the index $j$. This index determines the value of Casimir
operator of $SL(2,R)$. The $(x,\bar x)$ coordinates can be regarded
as the boundary coordinates of the $SL(2,R)$ target space-time, and
$(w,\bar w)$ are worldsheet coordinates. One can replace the
boundary coordinates with the numbers $(m,\bar m)$, defined via
transformation
\eqn\xbxmbmtr{V _{j;m,\bar m}(w,\bar w)=\int d^2xx^{j+m}\bar
x^{j+\bar m}V_j(x,\bar x;w,\bar w)\,.}
OPE of $SL(2,R)$ currents and $SL(2,R)$ primaries are \foot{For
Euclidean $SL(2,R)$,
\eqn\JVopeEucl{J^3(z)V_{j;m,\bar m}(w,\bar
w)=\frac{im}{z-w}V_{j;m,\bar m}(w,\bar w) +\cdots\,.}
Therefore
\eqn\els{\eta_{33}(J^3)^2(z)V_{j;m,\bar m}(w,\bar
w)=-\frac{m^2}{z-w}V_{j;m,\bar m}(w,\bar w)}
is true for both Euclidean and Minkowski signatures.}
\eqn\JVope{J^3(z)V_{j;m,\bar m}(w,\bar w){=}\frac{m}{z-w}V_{j;m,\bar
m}(w,\bar w){+}\cdots\,,\;\; J^\pm (z)V_{j;m,\bar m}(w,\bar w){=}
\frac{m\mp j}{z-w}V_{j;m\pm 1,\bar m}(w,\bar w){+}\cdots\,.}
From this one finds how the $SL(2,R)$ currents act on the primaries:
\eqn\SLtrtrp{J^3_0\cdot V_{j;m,\bar m}(w,\bar w)=mV_{j;m,\bar
m}(w,\bar w)\,,\quad J^\pm _0\cdot V_{j;m,\bar m}(w,\bar w)= (m\mp
j)V_{j;m\pm 1,\bar m}(w,\bar w)\,,}
with all other $J_n^A\cdot V_{j;m,\bar m}(w,\bar w)=0$, $n\geq 1$.

Second order $SL(2,R)$ Casimir operator is given by
\eqn\Ctdef{C_2=\eta_{AB}J_0^AJ_0^B\equiv
-(J_0^3)^2+\frac{1}{2}\{J_0^+,J_0^-\}\,.}
Here
\eqn\Jnot{J_0^1=\frac{1}{2}(J_0^++J_0^-)\,,\quad\quad
J_0^2=\frac{i}{2}(J_0^--J_0^+)\,.}
It takes place
\eqn\CtV{C_2\cdot V _j(w,\bar w)=-j(j+1)V_j(w,\bar w)\,.}
This expression is also true for Euclidean $SL(2,R)$, due to
\JVopeEucl. Then clearly for $SL(2,R)$ algebra with currents of
weight $\hat k$,
\eqn\LzactV{L_0\cdot V _j(w,\bar w)=-\frac{j(j+1)}{\hat k-2}V
_j(w,\bar w)\,,}
which gives the conformal dimension of $V _j$
\eqn\Deltaj{\Delta _j=-\frac{j(j+1)}{\hat k-2}\,.}

In the superstring theory one considers the total bosonic currents
$J^a$, which include contributions from worldsheet fermions, dual to
$SL(2,R)$ currents. The level of total $SL(2,R)$ currents is equal
to $\hat k+2$, if $\hat k$ denotes the level of purely bosonic
currentt, and therefore the conformal dimension of the $V_{jm\bar
m}$ is equal to $\Delta _j=-\frac{j(j+1)}{\hat k}$.

\subsec{Gauged WZW model on the $\frac{SL(2,R)\times U(1)}{U(1)}$}

Let us review the derivation \GiveonGE\ of the gWZW action on the
$\frac{SL(2,R)\times U(1)}{U(1)}$ coset.

Perform the following asymmetric gauging of the $U(1)$ subgroup of
$SL(2,R)\times U(1)$ group with the parameter $\tau$:
\eqn\gaugedef{\left(g,\; x_L,\;
x_R\right)\sim\left(e^{\tau\cos\psi\sigma_3/\sqrt{\hat k}}g
e^{\tau\sigma_3/\sqrt{\hat k}},\; x_L+\tau\sin\psi,\; x_R\right)\,.}
The condition that the gauge transformation leaves the action
invariant is
\eqn\gaugeinv{{\rm Tr}\,\left( T_L^2- T_R^2\right)=0\,,}
where $T_{L}$ and $T_R$ are the generators of left-moving and
right-moving sectors of the gauged $U(1)$ group.

Let us write the element of the $SL(2,R)\times U(1)$ group as
\eqn\Gdefg{G=\left({g\atop 0}\;{0\atop \exp\left(\sqrt{\frac{2}{\hat
k}}\,x\right)}\right)}
Then $G$ is a field of the $SL(2,R)\times U(1)$ WZW model at level
$\hat k$:
\eqn\Gact{S[G]=\frac{\hat k}{4\pi}[\int d^2z{\rm Tr}(G^{-1}\p
GG^{-1}\bar \p G)-\frac{1}{3}\int _B{\rm Tr}(G^{-1}dG)^3]=}
\eqn\GactT{=\frac{\hat k}{4\pi}[\int d^2z{\rm Tr}(g^{-1}\p
gg^{-1}\bar \p g)-\frac{1}{3}\int _B{\rm
Tr}(g^{-1}dg)^3]+\frac{1}{2\pi}\int d^2z\p x \bar\p x\,.}
The gauge transformation \gaugedef\ acts on $G$-field as
\eqn\gGtr{G\rightarrow e^{T_L\tau}Ge^{T_R\tau}\,}
where the generators of left and right sectors of the $u(1)$ algebra
are
\eqn\TLTRdef{T_L=\left({\frac{1}{\sqrt{\hat k}}\cos\psi\sigma^3\atop
0}\;{0\atop \sqrt{\frac{2}{\hat k}}\sin\psi}\right)\,,\quad
T_R=\left({\frac{1}{\sqrt{\hat k}}\sigma^3\atop 0}\;{0\atop
0}\right)}
These generators satisfy anomaly-free condition \gaugeinv. Because
of this condition is satisfied we can make gauge fields
non-dynamical, as it is shown bellow.

Consider compensator fields (gauge field `prepotentials'):
\eqn\UVdef{U=\exp\left(-u T_L\right)\,,\quad\quad V=\exp\left(-v
T_R\right)\,.}
Define gauge transformation of compensator fields as
\eqn\compgtr{u\rightarrow u+\tau\,,\quad\quad v\rightarrow v+\tau}
The combination $UGV$ is clearly invariant under gauge
transformations \gaugedef, and therefore the WZW-action $S[UGV]$ is
gauge-invariant.

But it contains terms which are quadratic in derivatives of
compensator field $u$ and quadratic in derivatives of compensator
field $v$. Such terms make the compensator (gauge) d.o.f. dynamical,
and therefore the theory with the action $S[UGV]$ instead of gauging
some degrees of freedom away adds more degrees of freedom.

Therefore let us consider instead the gWZW action
\eqn\gwzwuc{S_g=S[UGV]-\frac{1}{2\pi}\int d^2z\p w\bar\p w\,,}
where we have introduced gauge-invariant field
\eqn\wdef{w=u-v\,.}

Due to the Polyakov-Wiegmann indentity
\eqn\pwapguv{S[UGV]=S[G]+S[U]+S[V]+}
\eqn\pwapguvt{+\frac{\hat k}{2\pi}\int d^2z{\rm
Tr}\left[G^{-1}\bar\p G\p VV^{-1}+U^{-1}\bar\p U\p
GG^{-1}+U^{-1}\bar\p UG \p VV^{-1}G^{-1}\right]}
Here
\eqn\SuSV{S[U]=\frac{1}{2\pi}\int d^2z\p u\bar\p u\,,\quad
S[V]=\frac{1}{2\pi}\int d^2z\p v\bar\p v}
and therefore
\eqn\SuSVV{S[U]+S[V]-\frac{1}{2\pi}\int d^2z\p w\bar\p
w=\frac{1}{2\pi}\int d^2z (\p v\bar\p u+\p u\bar \p
v)=\frac{1}{\pi}\int d^2z A\bar A \,,}
where
\eqn\AbarAd{A=-\p v\,,\quad\quad\bar A=-\bar\p u\,.}
The action term \SuSVV\ is non-dynamical, as it is expected in gWZW
model with asymmetric gauging, satisfying anomaly-free condition
\gaugeinv. As a result, the gWZW action on the $\frac{SL(2,R)\times
U(1)}{U(1)}$ is given by
\eqn\gWZWtot{S_g=S[g]+\frac{1}{2\pi}\int d^2z\p x\bar \p x+}
\eqn\gWZWtott{+\frac{1}{2\pi}\int d^2z \left[A\sqrt{\hat k}{\rm
Tr}\, (g^{-1}\bar\p g\sigma ^3)+\bar A\left(\sqrt{\hat k}{\rm Tr}(\p
gg^{-1}\sigma ^3)\cos\psi+2\sin\psi\p x\right)+\right.}
\eqn\gWZWtottt{\left.+A\bar A\left(2+{\rm Tr}(g^{-1}\sigma
^3g\sigma^3)\cos\psi\right)\right]\,.}

\appendix{B}{Heterotic gravity approximation}

In the type-II supergravity, considered in the section 5, two gauge
fields appear as $G_{x\mu}$ and $B_{x\mu}$ components after
Kaluza-Klein reduction of the compact $x$ coordinate. The
two-dimensional charged black hole is also a solution \McGuiganQP\
of heterotic supergravity equations of motion. In this section we
compute graviton and gauge field two-point functions in the
two-dimensional charged black hole background in heterotic
supergravity. In this case there is just one background gauge field.
We solve fluctuation equations of motion for the shear components of
graviton and the transverse component of the gauge potential and
find one hydrodynamic mode. Matching the obtained dispersion
relation with the result obtained in the study of thermodynamics of
the 2d charged black hole we derive shear viscosity to entropy ratio
for any value of~$\psi$.

The two-loop beta-functions of bosonic fields in heterotic string
theory are \refs{\CallanIA,\McGuiganQP}
\eqn\gmneq{\beta^G_{\mu\nu}=R_{\mu\nu}+2\nabla_\mu\p_\nu\Phi
-\frac{1}{2}g^{\lambda\rho}F_{\mu\rho}F_{\nu\lambda}\,,}
\eqn\dileq{\beta^\Phi=\frac{1}{4}F^2-R+c+4(\p\Phi)^2-4\nabla^2\Phi\,,}
\eqn\Amueq{\beta^A_\nu=g^{\mu\lambda}(\nabla _\mu F_{\nu\lambda}-2
F_{\nu\lambda}\p_\mu\Phi)\,.}
Corresponding equations of motion, $\beta^{G,B,\Phi}=0$, have the
$CBH\times R^{d-1}$ solution,
\eqn\solgr{g_{\mu\nu}={\rm diag}\{-f(r),1/f(r),1,...,1\}\,,\quad
f(r)=1-2Me^{-Qr}+q^2e^{-2Qr}}
\eqn\solgrt{\Phi=\Phi_0-\frac{Qr}{2}\,,\quad\quad
F_{tr}=F(r)=\sqrt{2}Qqe^{-Qr}\,.}
Here $Q=2/\sqrt{\hk}$.

Consider fluctuations $h_{\mu\nu}$, $a_\mu$ and $\varphi$ around
this solution. Use the diffeomorphism invariance to fix $h_{\mu
r}=0$. Among $d+1$ space-time coordinates we have $t,r$ coordinates
of CBH and $d-1$ flat coordinates. Let us consider $CBH\times R^2$.
Choose $X$ to be the $R^2$ direction of propagation of excitations
(with momentum $p$) and choose $Y$ to be the $R^2$ direction,
transverse to the direction of propagation of excitations.
Fluctuations depend on $t,r,X$. The dependence on $t$ and $X$ in
momentum representation boils down to the factor $e^{-i\omega
t+ipX}$.

Plugging $g_{\mu\nu}+h_{\mu\nu}$, $A_\mu+a_\mu$ and $\Phi+\varphi$
with the most general fluctuations, we obtain shear channel
expressions (prime denotes differentiation w.r.t. $r$)
\eqn\betarY{\beta_{rY}^G=\frac{ie^{-i\omega t+ipX}}{2f}(\omega
h_{tY}'+pfh_{XY}'-\omega Fa_Y)}
\eqn\betaXY{\beta_{XY}^G=e^{-i\omega
t+ipX}\left(-\frac{1}{2f}(\omega
ph_{tY}+\omega^2h_{XY}+ff'h_{XY}'+f^2h_{XY}'')+f\Phi'h_{XY}'\right)}
\eqn\betatY{\beta_{tY}^G=\frac{e^{-i\omega
t+ipX}}{2}\left(p^2h_{tY}+\omega
ph_{XY}-fh_{tY}''+2f\Phi'h_{tY}'+fFa_Y'\right)}

The equations of motion in shear channel are therefore
\eqn\eqrY{\omega h_{tY}'+pfh_{XY}'-\omega Fa_Y=0}
\eqn\eqXY{\omega
ph_{tY}+\omega^2h_{XY}+ff'h_{XY}'+f^2h_{XY}''-2f^2\Phi'h_{XY}'=0}
\eqn\eqtY{p^2h_{tY}+\omega
ph_{XY}-fh_{tY}''+2f\Phi'h_{tY}'+fFa_Y'=0}

Consider diff-invariant field
\eqn\ginvZ{Z=\omega h_{XY}+ph_{tY}}
Using \ginvZ\ and \eqrY\ express
\eqn\hZ{h_{tY}'=\frac{\omega ^2Fa_Y-pfZ'}{\omega^2-p^2f}\,,\quad
h_{XY}'=\omega\frac{Z'-pFa_Y}{\omega^2-p^2f}}

The equations \eqXY\ and \eqtY\ after one substitutes \hZ\ into
them, both give rise to the same equation (due to $F'-2F\Phi'=0$)
\eqn\Zeq{Z''+\left(\frac{\omega^2f'}{f(\omega^2-p^2f)}-2\Phi'\right)Z'+
\frac{\omega^2-p^2f}{f^2}Z-\frac{pF}{f}\left(fa_Y'+\frac{\omega^2
f'}{\omega^2-p^2f}a_Y\right)=0}

Compute the beta-function for gauge field fluctuation $a_Y$ (choose
the gauge $a_r=0$)
\eqn\betaAY{\beta^A_Y=-e^{-i\omega
t+ipX}f\left(a_Y''+\frac{f'-2f\Phi'}{f}a_Y'+\frac
{\omega^2-p^2f}{f^2}a_Y-\frac{Fh_{tY}'}{f}\right)}
Express $h_{tY}'$ using \hZ. The equation on $a_Y$ is then
\eqn\aYeq{a_Y''+\frac{f'-2f\Phi'}{f}a_Y'+\left(\frac
{\omega^2-p^2f}{f^2}-\frac{\omega^2
F^2}{f(\omega^2-p^2f)}\right)a_Y+\frac{pFZ'}{\omega^2-p^2f}=0}

Before proceeding, rescale
\eqn\Qres{\r =rQ\,,\quad\w=\omega/Q\,,\quad \pp=p/Q\,,\quad \Z=Z/Q}
The dependence on $Q$ disappears from both fluctuation equations,
and due to \solgrt\ we obtain
\eqn\fPhi{f=1-2Me^{-\r}+q^2e^{-2\r}\,,\quad\Phi=\Phi_0-\frac{\r}{2}}
Introduce new radial coordinate $u=e^\r$. Then inner and outer
horizons are located at
\eqn\upm{u_\pm =M\pm\sqrt{M^2-q^2}}
The equations of motion become (substitute $q=\sqrt{u_+u_-}$)
\eqn\aYu{\frac{d^2a}{du^2}+\left(\frac{1}{u-u_-}+\frac{1}{u-u_+}\right)
\frac{da}{du}+\frac{1}{(u-u_-)(u-u_+)}\left(\frac{\w^2u^2-\pp^2(u-u_-)(u-u_+)}{(u-u_-)(u-u_+)}-\right.}
\eqn\aYu{-\left.\frac{2u_+u_-\w^2}{\w^2u^2-\pp^2(u-u_-)(u-u_+)}\right)a+
\frac{\pp\sqrt{2u_+u_-}}{\w^2
u^2-\pp^2(u-u_-)(u-u_+)}\frac{d\Z}{du}=0}
\eqn\Zu{\frac{d^2\Z}{du^2}+\frac{1}{u}\left(2+\frac{\w^2 u^2}{\w^2
u^2-\pp^2(u-u_-)(u-u_+)}\left(\frac{u_+}{u-u_+}+\frac{u_-}{u-u_-}\right)\right)
\frac{d\Z}{du}-\frac{\pp\sqrt{2u_+u_-}}{u^2}\frac{da}{du}+}
\eqn\Zu{+\frac{\w^2u^2{-}\pp^2(u-u_-)(u-u_+)}{(u-u_-)^2(u-u_+)^2}\Z{-}\frac{\pp\sqrt{2u_+u_-}}{u}
\frac{\w^2}{\w^2u^2{-}\pp^2(u-u_-)(u-u_+)}\left(\frac{u_+}{u-u_+}{+}
\frac{u_-}{u-u_-}\right)a{=}0}

In the near horizon limit $v=u-u_+\ll 1$ equations \aYu\ and \Zu\
give rise to
\eqn\qYunh{\frac{d^2a}{dv^2}+\frac{1}{v}\frac{da}{dv}+\frac{\w^2u_+^2}
{(u_+-u_-)^2v^2}a=0}
\eqn\Zunh{\frac{d^2\Z}{dv^2}+\frac{1}{v}\frac{d\Z}{dv}+\frac{\w^2u_+^2}
{(u_+-u_-)^2v^2}\Z=0}
The incoming-wave solutions are
\eqn\aYuincw{a_Y(u)=C_1(u-u_+)^\frac{-i\w u_+}{u_+-u_-}\,,\quad
\Z(u)=C_2(u-u_+)^\frac{-i\w u_+}{u_+-u_-}}

In the asymptotic region $u\gg 1$ equations \aYu\ and \Zu\ give rise
to
\eqn\aYnb{\frac{d^2a}{du^2}+\frac{2}{u}\frac{da}{du}+\frac{\w^2-\pp^2}{u^2}a=0}
\eqn\ZYnb{\frac{d^2\Z}{du^2}+\frac{2}{u}\frac{d\Z}{du}+\frac{\w^2-\pp^2}{u^2}\Z=0}
with the solution
\eqn\aYnbsol{a_Y=\AA _a
u^{\frac{1}{2}(-1+\sqrt{1+4(\pp^2-\w^2)})}+\BB_a
u^{\frac{1}{2}(-1-\sqrt{1+4(\pp^2-\w^2)})}}
\eqn\ZYnbsol{\Z=\AA_Z
u^{\frac{1}{2}(-1+\sqrt{1+4(\pp^2-\w^2)})}+\BB_Z
u^{\frac{1}{2}(-1-\sqrt{1+4(\pp^2-\w^2)})}}

We solve numerically the equations \aYu\ \Zu\ with boundary
conditions \aYuincw. Zeroes of determinant of the leading behavior
coefficients matrix
\eqn\zerodet{\AA _Z^{(1)}\AA _a^{(2)}-\AA _Z^{(2)}\AA _a^{(1)}}
are located at
\eqn\zdet{\w=-i\pp^2\cos^2(\psi/2)}
Due to \Qres\ and $Q=2/\sqrt{\hk}$ from \zdet\ it follows
\eqn\shzerPth{ \omega =- i\frac{\sqrt{\hat
k}\cos^2(\psi/2)}{2}p^2\,,}

Finally, matching the dispersion relation \shzerPth\ to the
dispersion relation \shzerPt, obtained in Section 2 from the
consideration of thermodynamics of the 2d charged black hole, we
conclude
\eqn\etas{\frac{\eta}{s}=\frac{1}{4\pi}}
is valid for any value of $\psi$, and due to $q=M\sin\psi$ it is
valid for any charge density.

\footatend\vfill\supereject\immediate\closeout\rfile\writestoppt
\baselineskip=14pt\centerline{{\bf References}}\bigskip{\frenchspacing%
\parindent=20pt\escapechar=` \input refs.tmp\vfill\eject}\nonfrenchspacing

\end